\documentclass[sigconf]{acmart}

\setcopyright{none}
\renewcommand\footnotetextcopyrightpermission[1]{}

\usepackage{listings}
\usepackage{xcolor}
\usepackage{colortbl}
\usepackage{algorithm}
\usepackage{algpseudocode}

\definecolor{codegreen}{rgb}{0,0.6,0}
\definecolor{codegray}{rgb}{0.5,0.5,0.5}
\definecolor{codepurple}{rgb}{0.58,0,0.82}
\definecolor{backcolour}{rgb}{0.95,0.95,0.92}

\AtBeginDocument{%
  \providecommand\BibTeX{{%
    \normalfont B\kern-0.5em{\scshape i\kern-0.25em b}\kern-0.8em\TeX}}}

\newif\ifsubmit
\submitfalse
\ifsubmit
    \newcommand{\stefanie}[1]{}
    \newcommand{\megan}[1]{}
    \newcommand{\mackenzie}[1]{}
    \newcommand{\faraz}[1]{}
    \newcommand{\liane}[1]{}
    \newcommand{\changes}[1]{}

\else

    \newcommand{\faraz}[1]{{\leavevmode\color[rgb]{0.0, 0.0, 0.0}{#1}}}

\fi

\usepackage{soul}

\usepackage{booktabs}
\usepackage{multirow}
\usepackage{graphicx}
\usepackage{microtype}

\begin{document}

\title[MiXR]{MiXR: Harvesting and Recomposing Real-World Geometry for In-Situ Fabrication}

\author{Demircan Tas}
\authornote{Both authors contributed equally to this research.}
\email{tasd@mit.edu}
\affiliation{%
  \institution{MIT CSAIL}
  \city{Cambridge}
  \state{MA}
  \country{USA}
}

\author{Faraz Faruqi}
\authornotemark[1]
\email{ffaruqi@mit.edu}
\affiliation{%
  \institution{MIT CSAIL}
  \city{Cambridge}
  \state{MA}
  \country{USA}
}

\author{Ning Zhang}
\email{ningrz@mit.edu}
\affiliation{%
  \institution{MIT CSAIL}
  \city{Cambridge}
  \state{MA}
  \country{USA}
}

\author{Arthur Caetano}
\email{caetano@ucsb.edu}
\orcid{0000-0003-0207-5471}
\affiliation{%
  \institution{UCSB}
  \city{Santa Barbara}
  \state{CA}
  \country{USA}
}

\author{Dingning Cao}
\email{dcao2028@mit.edu}
\affiliation{%
  \institution{MIT CSAIL}
  \city{Cambridge}
  \state{MA}
  \country{USA}
}

\author{Niccolò G. Meniconi}
\email{nic.meniconi@gmail.com}
\affiliation{%
  \institution{Arizona State University}
  \city{Tempe}
  \state{AZ}
  \country{USA}
}

\author{Oğuz Arslan}
\email{oguz.arslan1@std.bogazici.edu.tr}
\affiliation{%
  \institution{Boğaziçi University}
  \city{Istanbul}
  \country{Turkey}
}

\author{Misha Sra}
\email{sra@ucsb.edu}
\orcid{0000-0001-8154-8518}
\affiliation{%
  \institution{UCSB}
  \city{Santa Barbara}
  \state{CA}
  \country{USA}
}

\author{Ruofei Du}
\email{ruofei@google.com}
\affiliation{%
  \institution{Google XR}
  \city{San Francisco}
  \state{CA}
  \country{USA}
}

\author{Stefanie Mueller}
\email{stefmue@mit.edu}
\affiliation{%
  \institution{MIT CSAIL}
  \city{Cambridge}
  \state{MA}
  \country{USA}
}

\author{Mustafa Doga Dogan}
\email{doga@adobe.com}
\affiliation{%
  \institution{Adobe Research}
  \city{Basel}
  \country{Switzerland}
}

\renewcommand{\shortauthors}{Tas et al.}

\begin{abstract}

Digital fabrication increasingly begins with generative AI: text or image prompts can produce a plausible 3D model in seconds. However, these models offer little control over spatial structure and no guarantee that the result is printable, and geometry that is dimensionally or personally specific, a hand's grip shape, a pet's exact proportions, a surface bent to fit a particular use, is difficult to specify through prompting alone. We present MiXR, an AR system that harvests geometry from real-world objects and composes it, through direct 3D manipulation, into fabricable designs. Users capture physical objects as 3D proxy meshes, extract meaningful segments, and arrange them explicitly in space; a generative model then integrates the arrangement into a single coherent mesh by operating on structured latent representations rather than merging raw geometry. In a 100-sample stress test, MiXR's composition pipeline converts open, non-manifold segmented input, none of which is watertight, into watertight output in 86/100 cases, reducing boundary edges by 99.3\% on average. We demonstrate the pipeline on three 3D-printed, functional artifacts built from harvested, dimensionally-specific geometry that general-purpose asset libraries cannot provide. A controlled user study ($N=12$) further shows that novices using MiXR produce designs rated significantly closer to their intended target, and report greater perceived control, than with a generative composition baseline.

 \end{abstract}




\begin{teaserfigure}
\centering
\includegraphics[width=0.80\textwidth]{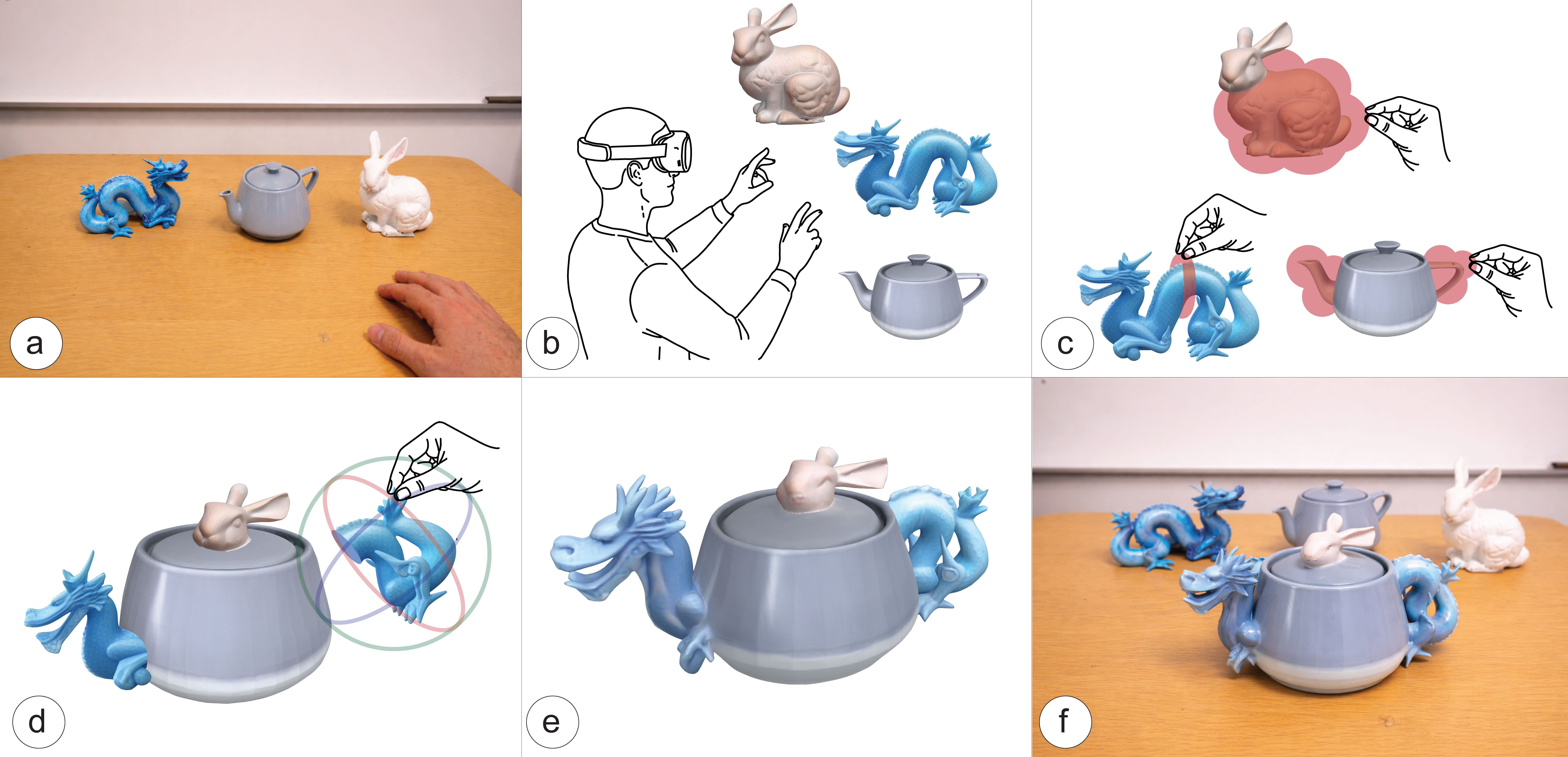}
\caption{MiXR enables in-situ compositional 3D modeling by harvesting geometry from physical objects and enabling users to precisely arrange them into a desired composition, which is then refined by generative AI into a unified 3D model. (a)~A user begins with everyday objects on a table. (b)~Using an XR headset, the user captures 3D models of the objects through hand gestures. (c)~Users select or remove regions of interest by painting over the specific part. (d)~Extracted segments are spatially arranged through direct 3D manipulation: translating, rotating, and scaling. (e)~A 3D Generative model refines the user's rough composition into a coherent unified 3D model. (f)~The final design is viewed in situ alongside the original objects.}
\label{fig:teaser}
\end{teaserfigure}

\maketitle

\section{Introduction}

Recent advances in 3D generative AI have made it possible for novice users to produce complex 3D models from simple text or image prompts~\cite{faruqi2024shaping, lin2023magic3d, trellis_2025}. Given a textual description (e.g., ``\textit{a headphone stand with a golden base}'') or a seed image, systems such as Magic3D~\cite{lin2023magic3d} and SAM3D~\cite{sam3d} can generate plausible 3D geometry in seconds, dramatically lowering the barrier to 3D content creation. However, these tools also constrain creative control~\cite{zamfirescu2023johnny, subramonyam2023bridging}. Most current systems operate as one-shot pipelines: users submit a prompt, receive one or several outputs, and have limited ability to iteratively adjust structure or preserve specific elements. Text prompts are often insufficient to fully capture spatial intent, particularly when a design idea involves precise geometric relationships such as a part rotated at a specific angle, a component scaled relative to another, or an asymmetric arrangement that is complex to verbalize. As a result, refinement becomes a trial-and-error process of re-prompting, where desired features may be lost with each regeneration, shifting creative practice from intentional design toward prompt optimization.

In contrast, creative practice across domains such as generative art, creative coding, and music production, relies on recombination, recontextualization, and iterative refinement. Prior studies demonstrate that non-experts struggle with underspecified prompts, limited feedback, and opaque model reasoning when interacting with generative AI systems~\cite{zamfirescu2023johnny, subramonyam2023bridging}. Recent HCI work has increasingly positioned generative models within human-in-the-loop workflows that expose intermediate representations and support localized intervention rather than one-shot generation~\cite{ petridis2023anglekindling, brade2023promptify, masson2024directgpt}. A key insight from this body of work is that creative engagement with generative systems emerges not from isolated prompts, but from structured interaction with intermediate artifacts.

Augmented reality provides a uniquely suited medium for this kind of compositional 3D design. Unlike screen-based modeling tools, AR allows users to manipulate virtual geometry directly in the physical space where a design will ultimately live. Users can see how components relate to real-world objects, adjust proportions relative to their environment, and work with spatial relationships through direct manual interaction rather than abstract parameter controls. Recent work in XR has demonstrated the potential of object-aware interaction~\cite{dogan2024augmented}, in-situ fabrication~\cite{weichel_mixfab_2014, arslan2025tinkerxr}, and real-time digital capture of physical artifacts~\cite{li2025interecon, hu2025thing2reality}. The vision of Programmable Reality~\cite{suzuki2025programmable} articulates how spatial computing and AI could enable computational manipulation of the physical world. However, these systems have primarily focused on augmentation, capture, or communication rather than enabling users to construct new geometry by composing parts across multiple objects.

Towards this goal, we present MiXR: an AR system for in-situ compositional 3D modeling that enables users to create new designs by harvesting geometry from their physical environment. Users capture real-world objects as 3D proxy meshes via SAM3D~\cite{sam3d}, extract semantically meaningful segments (e.g., handles, lids, structural elements), and recombine these segments through direct spatial manipulation in AR. Once a user has arranged segments into a coarse assembly that reflects their intended design, MiXR's latent-space composition pipeline synthesizes the composition into coherent, unified geometry. This hybrid workflow separates two concerns: users define spatial structure explicitly through direct manipulation, while the generative model handles the low-level geometric integration. Because the composed geometry is destined for physical fabrication rather than screen display, this integration step must do more than preserve spatial intent: it must also repair the open, non-manifold boundaries that segmentation introduces, since printing requires closed, manifold geometry that defines a clear interior volume~\cite{stress_relief}. The result is a level of compositional control and fabrication reliability that is difficult to achieve through prompting alone, particularly for designs involving multiple components with specific spatial relationships.

We characterize this fabrication pipeline through two technical evaluations. A controlled transform sweep quantifies how composition fidelity degrades under increasing spatial separation and scale mismatch between segments, and a 100-sample watertightness stress test shows that MiXR's decode stage converts open, non-manifold segmented input, none of which is watertight, into watertight and manifold meshes in 86\% of cases, a 99.3\% mean reduction in open boundary edges. We further demonstrate the pipeline end-to-end on three fabricated, functional artifacts: a personalized pet-shaped book support, a chair built from repurposed household objects, and an accessibility grip for an umbrella, each combining harvested, dimensionally-specific geometry that general-purpose asset libraries cannot easily provide.

To evaluate whether novices can reliably drive this fabrication pipeline, we conducted a controlled user study ($N=12$) comparing MiXR's segment-and-compose workflow against a generative composition baseline, where participants described the desired arrangement verbally and a VLM + 3D Generative model created the composed result. Both conditions used the same physical objects and 3D reconstruction backend, though the baseline additionally relies on a text-to-image generation stage that MiXR does not use. We tested across two qualitatively different tasks: an abstract building block arrangement and a freeform real-world object composition, to evaluate whether MiXR's advantage holds across different spatial reasoning demands. Our results show that MiXR yields significantly higher perceived control, spatial fidelity, and target matching across all measures, with participants unanimously preferring it for accuracy and control (all 12/12, $p < .001$). Participants experienced MiXR's hands-on effort as engaging rather than taxing, and 11 of 12 reported feeling more attached to designs they composed directly.

\noindent In summary, we make the following contributions:

\begin{itemize}

\item A compositional framework, operating on structured latent representations, that composes user-arranged, harvested geometry into unified, fabricable meshes without retraining, validated by a 100-sample watertightness stress test showing the pipeline converts open, segmented input into watertight and manifold geometry in 86\% of cases, a 99.3\% mean reduction in open boundary edges.

\item An end-to-end AR system integrating object capture, segmentation, spatial recomposition, and geometry refinement into a single in-situ fabrication workflow, demonstrated on three fabricated, functional artifacts built from dimensionally-specific, harvested geometry that general-purpose asset libraries cannot provide.

\item A controlled user study ($N=12$) showing that novices can reliably drive this fabrication pipeline through direct segment-level composition, producing designs closer to their intended target than VLM-mediated generative composition, across both abstract and real-world tasks.

\item Design guidelines for balancing user control and AI delegation in compositional fabrication tools for XR.

\end{itemize}

  \begin{figure}
    \centering
    \includegraphics[width=0.8\linewidth]{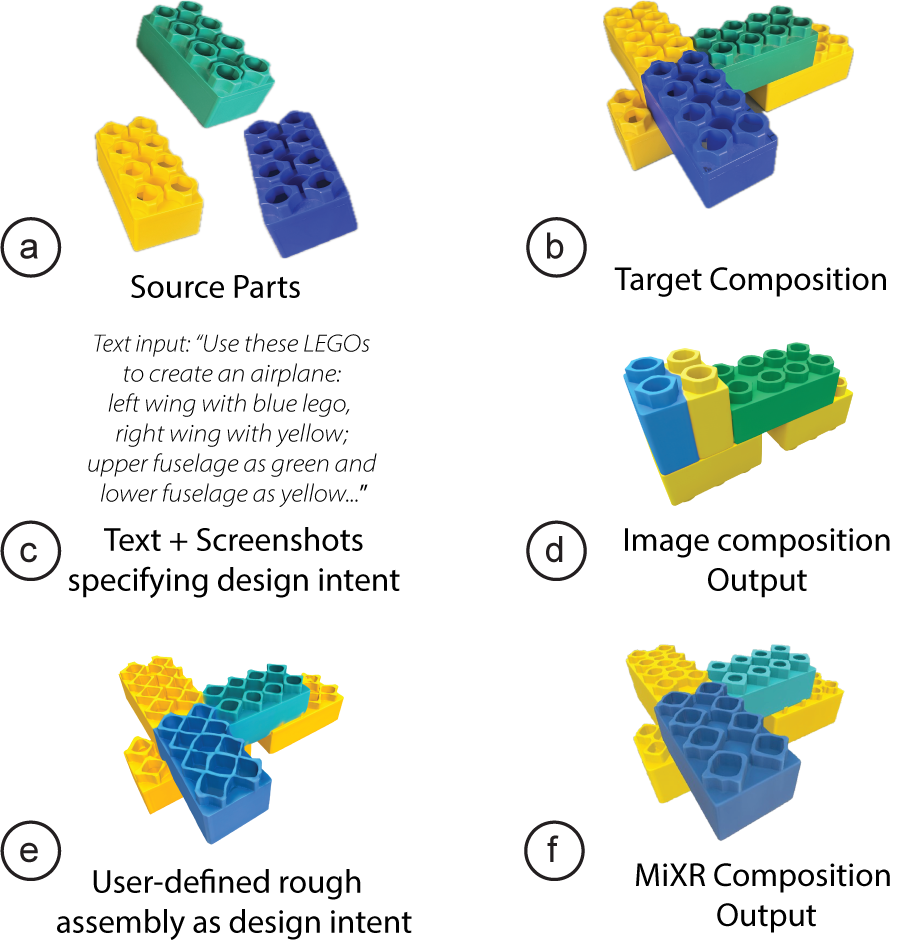}
    \caption{Compositional 3D modeling of a building block design. (a)~Source blocks. (b)~Target composition. (c–d)~Using the generative composition baseline, verbal prompts fail to create the intended design. (e–f)~With MiXR, the user arranges segments in 3D space (e), the system refines the assembly into a unified mesh that preserves the intended structure (f).}

    \label{fig:motivating_example}
\end{figure}

\subsection{Motivating Example}

Consider a simple compositional task: assembling five interlocking building blocks into a specific design (Fig~\ref{fig:motivating_example}). When provided with screenshots of the source parts and a detailed text description specifying which block should serve as each component, state-of-the-art VLM image generation model~\cite{gemini3report2025} produces outputs that fail to achieve the design intent of the user; building block pieces are modified and incorrectly placed (Figure~\ref{fig:motivating_example}d). This reflects a broader finding from prior work: users struggle to fully articulate spatial and compositional intent through text alone~\cite{zamfirescu2023johnny, subramonyam2023bridging}. MiXR is built on the core idea that a rough spatial assembly is a more reliable input than language as it directly encodes part placement, orientation, and relative scale, leaving less room for misinterpretation. In the same building task, a user's rough arrangement of the five bricks (Fig~\ref{fig:motivating_example}e) is encoded as a latent voxel structure~\cite{sam3d}, and then refined into a unified mesh that faithfully preserves the user's design intent (Fig~\ref{fig:motivating_example}f).

\section{Related Work}

We situate our work at the intersection of human-computer interaction, generative AI, augmented reality, and computational fabrication. Specifically, we draw on prior research in three areas: (1) advances in 3D generative models and compositional design tools, (2) human-in-the-loop customization of generative AI outputs, and (3) in-situ design and fabrication in extended reality.

\subsection{3D Model Generation and Design Tools}
Reconstructing 3D models from a single image or text description is a longstanding challenge in computer vision. Recent methods learn shape priors from large collections to infer geometry from minimal input~\cite{klokov2020discrete, chou2023diffusion, gao2022get3d}, and the LRM architecture~\cite{hong2023lrm} and its successors~\cite{xu2024instantmesh, tang2024lgm, boss2024sf3d} have demonstrated generalizable single-image 3D reconstruction. Multi-view diffusion with Gaussian-based representations have further improved fidelity and rendering efficiency~\cite{mu_gaussian, mvs_gaussian, Go_2025_CVPR}. TRELLIS~\cite{trellis_2025} introduced Structured Latents (SLATs) that jointly encode geometry and appearance, while SAM3D~\cite{sam3d} extended single-image reconstruction to robust in-the-wild performance. MiXR uses SAM3D as its reconstruction backbone and builds on the SLAT representation for compositional refinement.

In fabrication-oriented design, recent work has explored how generative models can be constrained to preserve functional properties: Style2Fab~\cite{faruqi2023style2fab} maintains functional affordances during personalization, TactStyle~\cite{faruqi2025tactstyle} controls tactile properties, and MechStyle~\cite{faruqi2025mechstyle} ensures structural viability through mechanical simulation. Most closely related to our technical evaluation, Speech to Reality~\cite{kyaw2025speech} discretizes and modifies AI-generated geometry to satisfy fabrication constraints, such as component connectivity, before robotic assembly; MiXR addresses an analogous gap for AR-composed geometry, showing that a latent-space decode step can restore watertightness to open, segmented input before 3D printing.

Generative 3D output is not, in general, ready to fabricate. Printing requires closed, manifold geometry that defines a clear interior volume~\cite{stress_relief}, and satisfying that topological requirement is separate from, and weaker than, the structural and process constraints that determine whether a printed artifact survives handling~\cite{stress_relief, stochastic_structural_analysis} or whether a given design is practically printable at all~\cite{dew2019producing}. Prior work narrows this gap either by constraining generation itself, through physically-aware generative networks~\cite{mezghanni2021physically} and differentiable simulation layers~\cite{mezghanni2022physical}, or by repairing geometry after the fact with general-purpose mesh processing tools~\cite{pymeshlab}. Standard repair treats the mesh purely as geometry: it closes boundaries and resolves non-manifold configurations without any representation of what the object is. MiXR's refinement step differs in kind rather than degree, because regenerating the composition through the generative model's own decoder completes geometry using a learned object prior rather than capping openings locally. We characterize the topological effect of this step in Section~\ref{sec:watertightness}; we do not benchmark it against dedicated repair tools, and we return to that comparison as future work.

Separately, HCI researchers have developed tools for part-based 3D modeling without generative AI. Meshmixer~\cite{schmidt2010meshmixer} supports mesh-library composition, PARTS~\cite{hofmann2018greater} and Attribit~\cite{chaudhuri2013attribit} enable design-intent-driven reuse and replacement, Grafter~\cite{roumen2018grafter} and AutoConnect~\cite{koyama2015autoconnect} create connectors and remix mechanisms, and parametric systems~\cite{shugrina2015fab, schulz2014design, veuskens2020coda} provide constrained customization. These tools are limited to static mesh libraries or parametric operations. MiXR extends this paradigm to in-situ part-based design with generative refinement.

\subsection{Human-in-the-Loop Customization with Generative Models}
While generative AI has expanded creative possibilities across domains, one-shot prompting remains a significant bottleneck. Research on novice interactions with AI systems~\cite{zamfirescu2023johnny, subramonyam2023bridging} consistently finds that users struggle to articulate intent through text alone and receive limited visibility into model reasoning. This has motivated a growing body of HCI work that positions generative models within interactive workflows.

Across creative domains, systems have introduced mechanisms for localized steering, iterative refinement, and structured reuse of generative outputs. In visual design, tools like PromptPaint~\cite{chung2023promptpaint}, Promptify~\cite{brade2023promptify}, and PromptCharm~\cite{wang2024promptcharm} enable spatially localized control over image generation, while DirectGPT~\cite{masson2024directgpt} extends direct manipulation principles to LLM-mediated content. In other domains, WorldSmith~\cite{dang2023worldsmith} and Spellburst~\cite{angert2023spellburst} support iterative forking and graph-based composition of generative pipelines. HistoryPalette~\cite{benharrak2025historypalette} and DreamSheets~\cite{almeda2024prompting} further demonstrate that capturing and reusing past generations strengthens creative engagement, a pattern echoed in studies of remixing in design practice~\cite{lee2010designing, subbaraman2023forking}.

A consistent finding emerges across this work: giving users explicit control over how generative outputs are preserved, modified, and recombined leads to greater alignment with intent. MiXR extends this principle to 3D composition for physical fabrication, where the intermediate artifact is a spatial assembly of segments, rather than a prompt history or attention map, that must ultimately resolve into a single fabricable mesh.

\subsection{In-Situ Design and Fabrication in AR}
XR technologies have increasingly been applied to situate design within the user's physical environment~\cite{dogan_fabricate_2022}. The vision of Programmable Reality~\cite{suzuki2025programmable} articulates how spatial computing and AI could enable computational manipulation of the physical world through digital twins and semantic understanding.

Several systems have explored immersive 3D design for novices. MixFab~\cite{weichel_mixfab_2014} demonstrated mixed-reality personal fabrication through direct interaction with virtual objects. TinkerXR~\cite{arslan2025tinkerxr} introduced in-situ CSG modeling with an integrated 3D printing workflow, pARam~\cite{stemasov_param_2024} provides in-situ parametric design through hand interactions, BrickStARt~\cite{stemasov_brickstart_2023} integrates tangible construction blocks with mixed reality, and Y-AR~\cite{feng2025yar} supports mixed-reality CAD for wire bending, evaluated through a novice usability study in which all participants fabricated a functional object. These tools demonstrate the value of situated design but rely on primitive shapes, parametric operations, or sketch-based modeling.

Recent work has also explored how physical context can enrich AI-mediated XR workflows. XR-Objects~\cite{dogan2024augmented} introduced object-centric AR menus driven by multimodal LLM reasoning. InteRecon~\cite{li2025interecon} extended digital capture to preserve the interactivity of personal artifacts. Thing2Reality~\cite{hu2025thing2reality} enabled spontaneous materialization of physical and digital objects in immersive environments. VRCopilot~\cite{zhang2024vrcopilot} integrates generative AI into VR authoring through intermediate wireframe representations. While these systems demonstrate the power of object-aware XR interaction, they primarily focus on augmentation, capture, or communication rather than enabling users to construct new geometry from composed parts.

MiXR contributes to this body of work by enabling users to harvest geometry from their environment, extract segments by design intent, recompose them through direct manipulation, and refine the assembly into unified geometry via a generative model.
\section{System Design}

MiXR is an in-situ 3D design system that enables users to construct new geometry by reusing existing objects as design material. Rather than treating AI-generated models as terminal outputs, the system supports segment-level harvesting and recomposition directly within the user's physical environment.

The system is structured around a three-mode workflow: \textit{Generate}, \textit{Segment}, and \textit{Compose}, presented through a unified spatial interface in augmented reality. Users move fluidly between modes: generating proxy 3D models from physical objects, painting regions to define segments by design intent, arranging segments through continuous direct manipulation, and invoking AI-based geometric refinement to produce coherent unified models. Users can move backward through these modes at any time, returning from Compose to Segment to adjust selections, or from Segment to Generate to capture additional objects. This reflects the iterative, non-linear nature of creative workflows.

Figure~\ref{fig:teaser} illustrates the end-to-end pipeline. In the following subsections, we describe each stage of the workflow in detail.

\subsection{Model Harvesting}

\begin{figure}
    \centering
    \includegraphics[width=0.8\linewidth]{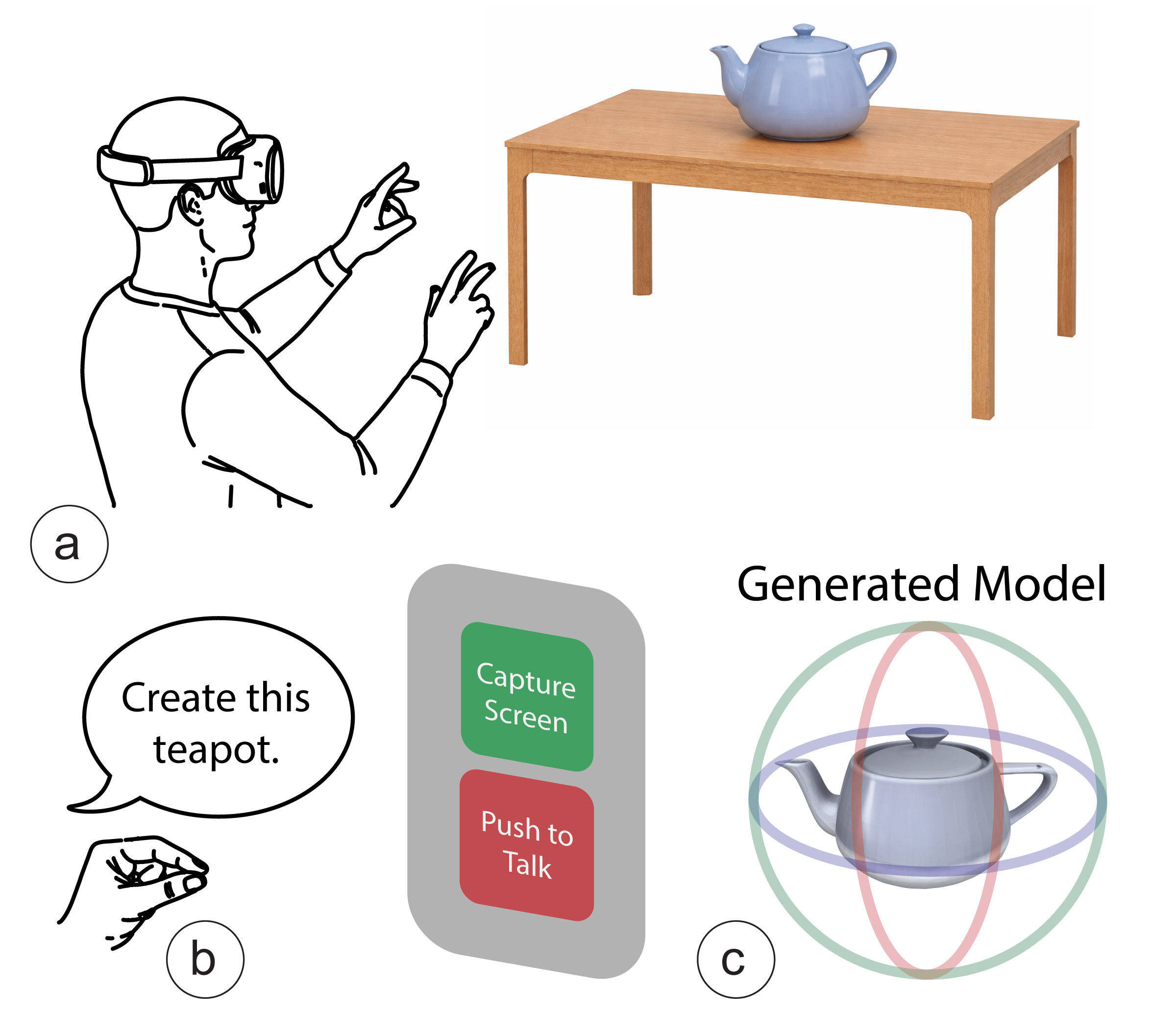}
\caption{MiXR allows users to harvest 3D models from physical objects in their environment. (a)~The user captures a screenshot of the target object with AR passthrough. (b)~A spoken prompt describes the intended object. (c)~The system generates the 3D model and returns the 3D mesh placed into the AR scene.}

    \label{fig:generation}
\end{figure}

In the first stage, users replicate (harvest) 3D objects from their environment they wish to use for downstream modeling. Users begin in \textit{Generate} mode, where they capture real-world objects in their environment as 3D proxy meshes (Fig.~\ref{fig:generation}). To replicate an object, the user captures a screenshot of the scene through the headset's passthrough cameras and provides a verbal description via push-to-talk speech input. The screenshot provides visual context while the prompt specifies which object to extract from the scene. The system sends both inputs to the SAM3D~\cite{sam3d} backend, which reconstructs a 3D mesh of the specified object and returns it as a glTF asset. The generated model is placed into the AR scene, where it can be repositioned and inspected.

Importantly, the server stores the generated glTF mesh together with its corresponding intermediate sparse latent volume (used by the decoder) and asset metadata, enabling the system to reuse the object as compositional material rather than treating generation as a terminal output.

Users may generate multiple assets within a single session, capturing different objects from their environment to build a library of 3D objects for subsequent segmentation and recomposition. Each generated asset is tracked as an independent workspace item that can be browsed, selected, duplicated, or deleted.

\subsection{Paint-Based Segmentation}

Once proxy models have been generated, users transition to \textit{Segment} mode to define which regions of each model to preserve for recomposition. Rather than relying on automated semantic segmentation (e.g., SAM~\cite{kirillov2023segment}), MiXR uses a manual paint-based selection approach. This design decision reflects a key distinction: semantic segmentation identifies parts that are meaningful to the model (e.g., ``handle,''), but these boundaries may not align with the user's design intent. A user might want the upper third of a vase regardless of semantic part boundaries, or a decorative band that spans multiple regions. By giving users direct control over segment boundaries, MiXR ensures that the segmentation reflects design intent rather than automated part classification.

\begin{figure}
    \centering
    \includegraphics[width=0.8\linewidth]{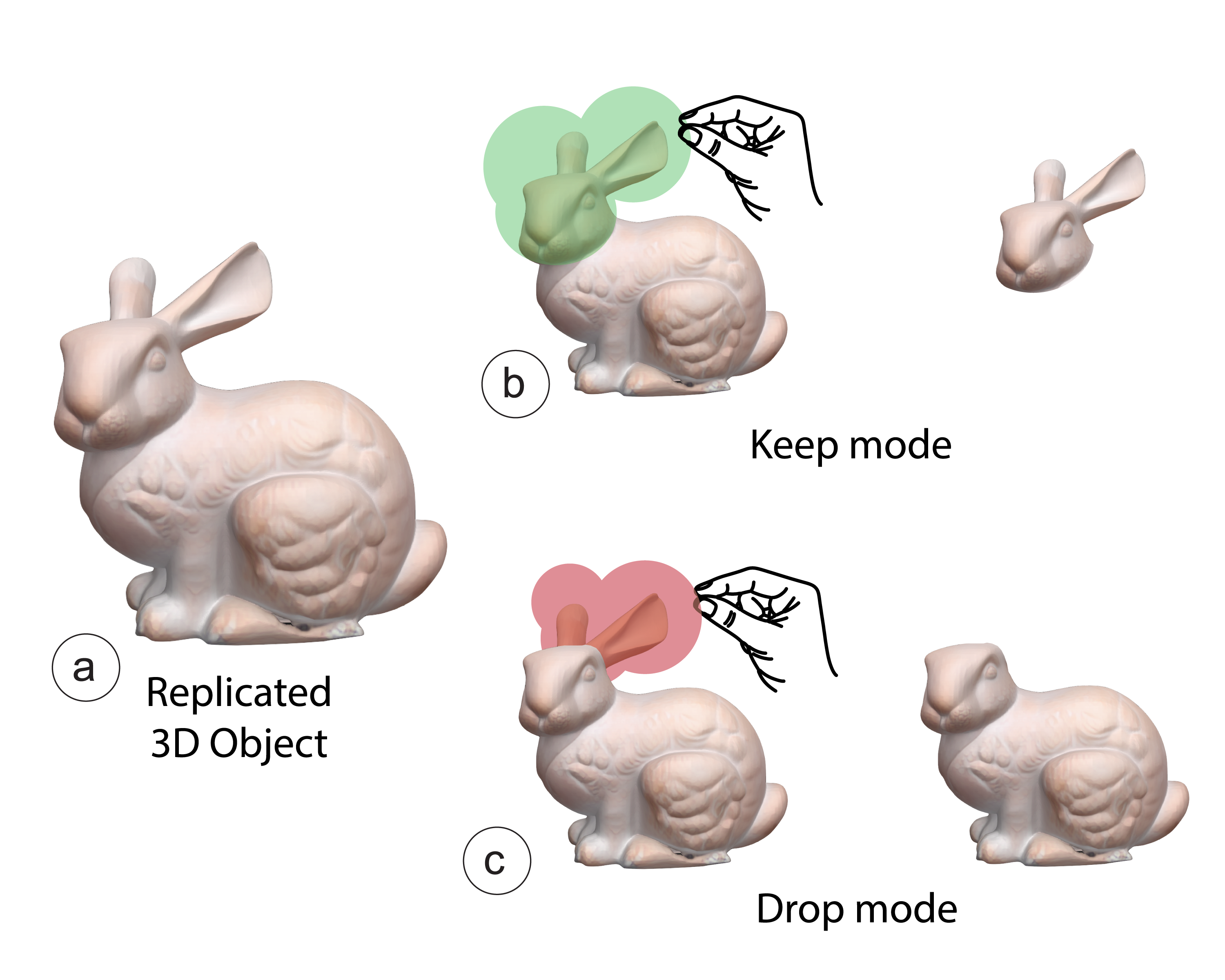}
  \caption{Paint-based segmentation. (a)~The captured 3D proxy mesh. (b)~In Keep mode (green), the user paints regions to preserve. (c)~In Drop mode (red), the user paints regions to remove. Users toggle between modes and adjust brush radius to refine boundaries.}

    \label{fig:segmentation}
\end{figure}

The segmentation tool operates as a 3D paint brush in AR (Fig.~\ref{fig:segmentation}). Users paint directly onto the surface of a proxy model using their hand; the system continuously samples the stroke path and marks each vertex within the brush radius according to the current paint mode. Two complementary modes are provided: \textbf{Keep} (green) marks vertices for preservation, and \textbf{Drop} (red) marks vertices for removal. Users can toggle freely between modes and adjust the brush radius to balance precision and coverage. Each vertex carries a boolean selection mask that determines whether it is included in the final segment. Because the brush operates in the mesh's local coordinate space with radius scaling that accounts for world-space scale, the interaction remains consistent regardless of how the model has been resized or repositioned. Users can segment multiple assets independently, building a collection of user-defined parts sourced from different objects.

\subsection{Spatial Recomposition}

\begin{figure}
    \centering
    \includegraphics[width=\linewidth]{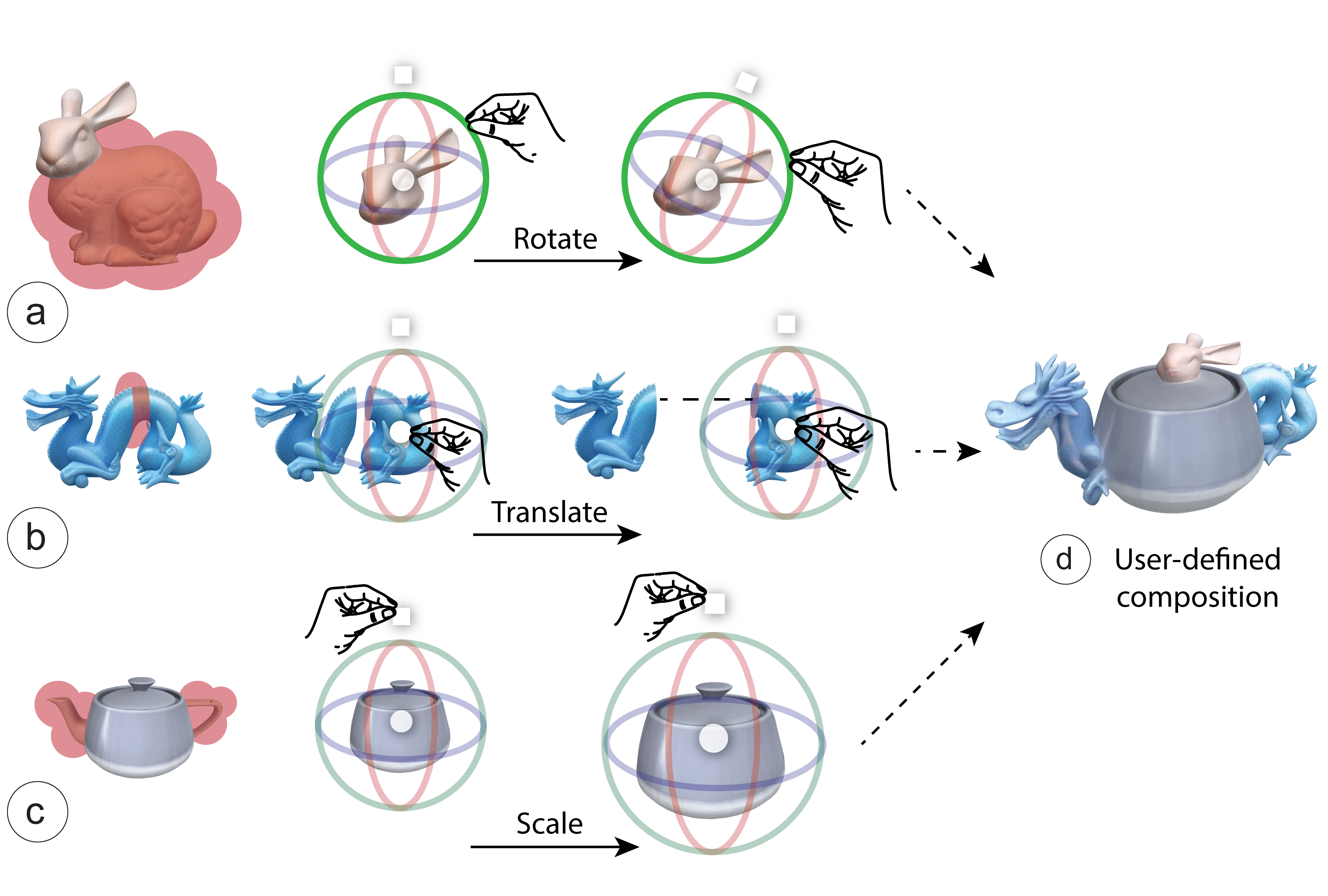}
    \caption{Spatial recomposition via direct manipulation. (a)~The segmented Stanford Bunny is rotated into position. (b)~The Stanford Dragon half is translated to form the handle. (c)~The Utah Teapot body is scaled to serve as the base. (d)~The resulting user-defined composition.}

    \label{fig:composition}
\end{figure}

In \textit{Compose} mode, users arrange their segmented assets into a spatial assembly that reflects the intended design (Fig.~\ref{fig:composition}). Each segment behaves as an independent 3D object that can be transformed through a custom gizmo system designed for direct hand interaction in AR. The transform gizmo provides axis-aligned rotation rings, a central translation sphere with 1:1 hand-displacement mapping, and a scale cube for uniform resizing, all using proximity-based direct grab detection.

Users can browse between assets, duplicate segments to reuse the same part multiple times in a composition, and delete assets that are no longer needed. As shown in Figure~\ref{fig:composition}, users iteratively adjust placement and proportions while previewing the result in context, ensuring that the emerging design fits both spatially and aesthetically within their physical environment.

\subsection{AI-Based Geometric Refinement}

After assembling a coarse composition, users invoke a refinement step to consolidate the arranged segments into coherent, unified geometry. MiXR's composition approach leverages the two-stage architecture of SAM3D~\cite{sam3d} by intervening between its intermediate representation and final decoding stages.

In SAM3D's standard pipeline, the Geometry Model first predicts a coarse shape as a $64^3$ voxel grid along with layout parameters (rotation, translation, scale). The Texture \& Refinement Model then extracts active voxels from this coarse shape and refines them using a sparse latent flow transformer, producing Structured Latent (SLAT) representations that encode fine geometric detail and surface appearance. These SLATs are decoded by a learned decoder into a final mesh.

 MiXR's composition pipeline operates on these intermediate representations in three stages:

\paragraph{Transform and selection application} For each segment, the system retrieves the glTF mesh and its corresponding SLAT representation from the backend. The user's spatial transforms are stored as homogeneous coordinate matrices and applied to both the mesh geometry and SLAT voxels, ensuring consistency between the AR view and the refinement input. The vertex-level selection mask is propagated to source SLAT voxels via a proximity threshold, so only user-selected regions contribute to the composed volume.

\paragraph{Latent union} The transformed and filtered SLAT volumes are combined via union into a single representation. The combined surface is voxelized into a composed support grid, with each voxel assigned the latent feature vector of its nearest source voxel. This nearest-neighbor projection preserves per-segment appearance without cross-boundary blending. The resulting volume encodes both \textit{where} geometry should exist (from the user's spatial composition) and \textit{what} it should look like (from the original per-segment SLATs).

\paragraph{Decoding} The combined SLAT volume goes through SAM3D's learned decoder, which synthesizes a unified mesh. As the decoder operates on a single latent volume rather than stitching separate meshes, it naturally handles gap-filling, surface harmonization, and texture consistency across formerly separate parts.

This approach requires no retraining or fine-tuning of the generative model: the decoder does not distinguish between a SLAT volume from a single reconstruction and one assembled from multiple user-arranged segments, so the user's spatial decisions are preserved while the decoder handles geometric integration. Because decoding regenerates geometry from the model's learned prior rather than patching the raw mesh, this step also repairs the open, non-manifold boundaries that segmentation introduces, restoring the topological validity that downstream fabrication requires (Sec.~\ref{sec:watertightness}).

The refined model replaces the coarse assembly in the AR scene. Users can inspect the result and return to Segment or Compose mode to adjust inputs before re-invoking refinement, or export the final model as a glTF asset, converted to a textured OBJ mesh at slicer upload for 3D printing, or used directly for integration into other applications or continued use in AR.

\subsection{Implementation}
MiXR is a WebXR application built on XRBlocks~\cite{xrblocks}, running in Google Chrome on the Samsung Galaxy XR headset with no native installation required. The spatial interface provides draggable panels for mode-specific controls. Speech input uses the Web Speech API; screenshot capture uses the WebXR Camera Access API. The backend runs on a cloud server (NVIDIA A40, 40\,GB VRAM), with approximately 15-second end-to-end latency for both generation and composition.

\section{Technical Evaluation}
\label{sec:technical}
MiXR's composition pipeline operates by combining user-defined SLAT volumes and decoding them through SAM3D's decoder. Since this decoder was trained on single-object volumes, compositions that expand the spatial extent of the combined volume, through large translations, scale mismatches, or complex arrangements, may degrade output fidelity. We characterize this operating regime through a controlled benchmark that systematically varies the relative transform between two source objects.

\subsection{Composition Robustness Under Relative Transform}

We evaluate how faithfully MiXR's decoder preserves the user's intended spatial arrangement by systematically varying the relative transform between two source objects.

\paragraph{Protocol.}

\begin{figure*}[t]
    \centering
    \includegraphics[width=0.33\textwidth]{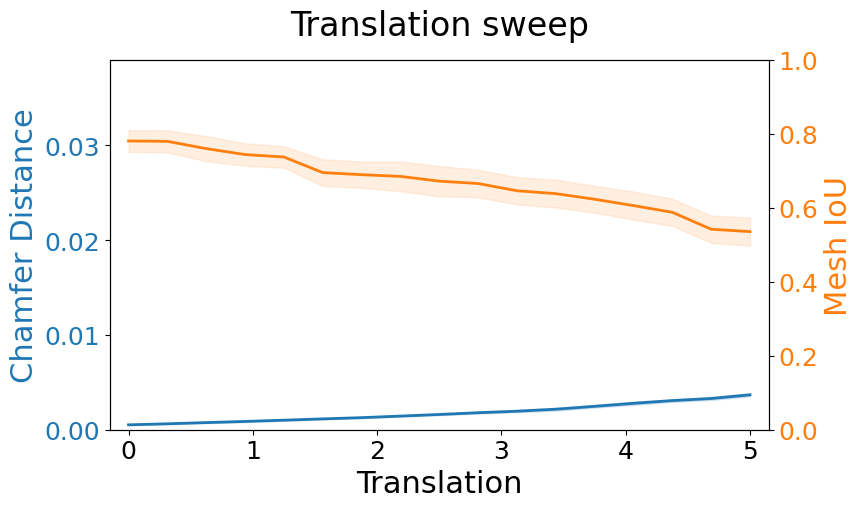}\hfill
    \includegraphics[width=0.33\textwidth]{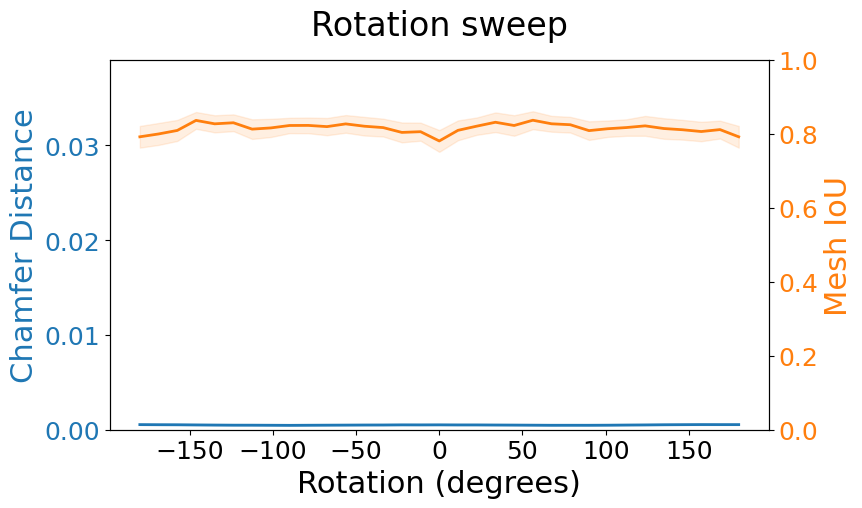}\hfill
    \includegraphics[width=0.33\textwidth]{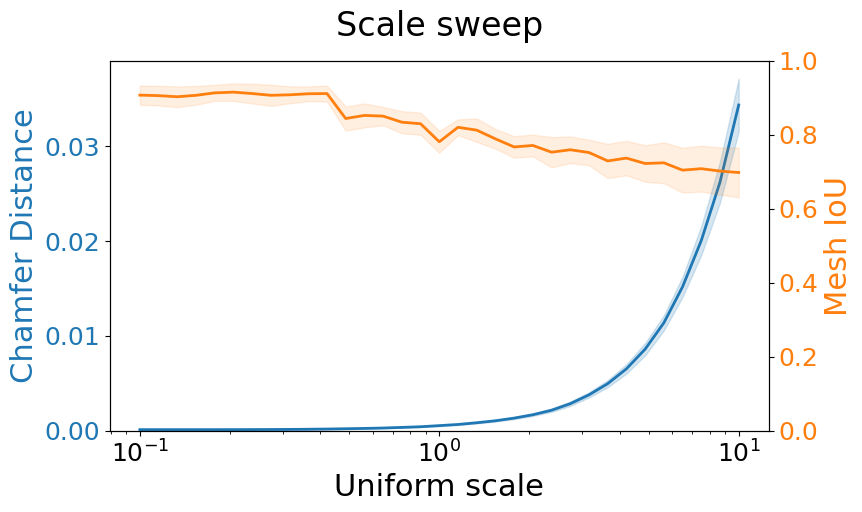}
    \caption{Mean Chamfer Distance (squared) and mean Mesh IoU across all 25 ordered category pairs for translation, rotation, and scale sweeps. Shaded bands show 95\% confidence intervals. Translation produces a monotonic degradation, rotation remains comparatively stable across the full sweep, and large uniform up-scaling causes a sharp nonlinear increase in error.}
    \label{fig:tech-eval-sweeps}
\end{figure*}

We select five 3D models from Objaverse~\cite{objaverse}'s most populous categories (antenna, banana, doughnut, ring, soccer ball), spanning elongated, toroidal, concave, and compact convex geometries, and evaluate all $25$ ordered pairs under three one-dimensional transform sweeps: translation along $x$ (linearly from $0.0$ to $5.0$ unit distance), rotation around $y$ ($-180^\circ$ to $180^\circ$), and uniform scaling (geometrically from $0.1\times$ to $10\times$), each with $33$ steps. All objects are normalized to unit bounding-box extent. For each step, the decoded mesh is compared against a \emph{reference composite}: the geometric union of the anchor and transformed meshes with no decoding, which captures distortion introduced by the latent regeneration. We report symmetric Chamfer Distance (squared) and Mesh IoU; shaded regions in Figure~\ref{fig:tech-eval-sweeps} show $95\%$ confidence intervals across all $25$ pairs. Per-pair breakdowns are provided in Appendix.

\paragraph{Baseline composition cost.}
Even at identity transforms (zero translation, zero rotation, unit scale), mean Mesh IoU is $0.78$ rather than $1.0$. This reflects the inherent reconstruction loss: the voxelization, and decoding processes in generative models are lossy operations for reconstruction. Thus, the baseline IoU represents the minimal loss with composition itself.

\paragraph{Translation.}
Translation produces the strongest monotonic degradation. The plotted values represent an \emph{additional} offset beyond a built-in baseline separation (approximately $1.6$--$2.0$ object extents), so objects are already non-overlapping at translation~$= 0$. Within the practical composition regime, (roughly $0$ to $2.5$ additional object extents) mean Mesh IoU declines smoothly from $0.78$ (95\% CI: $[0.75, 0.81]$) to $0.67$ ($[0.63, 0.71]$), while mean squared Chamfer rises from $5.2 \times 10^{-4}$ to $1.6 \times 10^{-3}$. The extended sweep to $10.0$ object extents is reported in Appendix.

\paragraph{Rotation.}
Rotation has minimal impact on composition fidelity. Across the full $-180^\circ$ to $180^\circ$ sweep, mean Mesh IoU remains between $0.78$ and $0.84$, while mean squared Chamfer stays between $4.6 \times 10^{-4}$ and $5.4 \times 10^{-4}$. Relative orientation does not meaningfully expand the spatial support of the composed voxel volume, so the decoder's fixed-resolution grid accommodates reorientation without the capacity pressure that translation and scaling introduce.

\paragraph{Scale.}
Uniform scaling produces a qualitatively different pattern. Above $1.0\times$, Chamfer error grows sharply and nonlinearly, reaching $3.8 \times 10^{-3}$ at $3.2\times$ and $3.4 \times 10^{-2}$ at $10\times$, while Mesh IoU declines more gradually (from $0.78$ to $0.70$). This divergence indicates that coarse occupancy overlap is partially preserved even as fine surface fidelity deteriorates rapidly, yielding a practical stability limit around $3$--$4\times$.

\paragraph{Interpretation.}
The three sweeps consistently characterize MiXR's operating regime: robust to reorientation, moderately sensitive to separation, and sensitive to dramatic scale increases. All three trends follow from a single architectural factor: the decoder reconstructs from a fixed-resolution $64^3$ voxel grid, so quality degrades when the combined SLAT volume's spatial extent exceeds what the grid can faithfully represent. Translation and up-scaling expand this extent, whereas rotation merely reorients content within a similarly sized volume. Notably, large translation errors at the sweep extremes are unlikely in practice, as users composing coherent designs will inherently keep components in close proximity.

\subsection{Topological Validity of Composed Meshes}
\label{sec:watertightness}

The evaluation above measures decoder distortion relative to a naive geometric union of source meshes; it does not test whether decoding produces topologically valid output, a prerequisite for fabrication. MiXR's paint-based segmentation removes faces from a source mesh without closing the resulting surface, so a segmented selection has open boundaries by construction. We test whether MiXR's decode stage repairs this topology or merely passes it through.

\paragraph{Protocol.}
We sampled 100 compositions from the $20$ non-self ordered category pairs (the same five Objaverse categories above) $\times$ $3$ translation axes $\times$ $5$ translation offsets, deterministic under a fixed random seed. For each sample, we kept the upper $65\%$ of a source mesh's vertices by height, simulating a paint-based selection, applied the sampled transform, and compared two conditions: the segmented meshes concatenated directly with no decode, and the same selections and transform passed through MiXR's composition pipeline (SLAT union followed by VAE decode). We measured watertightness, manifoldness, and boundary-edge count for each output with \texttt{trimesh}.

\paragraph{Results.}
Raw segmented-and-transformed concatenations were never watertight or manifold ($0/100$), with a mean of $3243$ boundary edges. Decoded compose outputs were watertight and manifold in $86/100$ cases, with a mean of $20$ boundary edges, a $99.3\%$ mean reduction in open boundaries. The decoder does not close every composition: the $14$ non-watertight cases concentrated at higher translation offsets ($7$ at the largest offset tested, $5$ at the next), and predominantly along the vertical axis ($10$ of $14$ cases), consistent with translation being the dominant stressor above. Even these failure cases substantially reduced boundary edges relative to the raw baseline, with the worst decoded case still an $83.67\%$ reduction. These results indicate the decode stage is not a cosmetic smoothing pass: it reconstructs substantially more topologically valid geometry from partial, spatially arranged, open-boundary input, output directly relevant to downstream fabrication.

\paragraph{Scope.}
This is a controlled synthetic segmentation protocol, not a reproduction of user-painted selections: retaining the upper $65\%$ of vertices by height approximates the open boundaries a real paint-based selection introduces, but does not cover the full range of selection shapes users can produce (e.g., disconnected regions, narrow appendages, or interior cavities). The $86\%$ watertightness result should be read as evidence that the decode stage substantially improves topological validity under this stress test, not as a guarantee that holds across all possible user selections.

\section{Applications}
\label{sec:applications}

We demonstrate MiXR's versatility through three application scenarios, each highlighting a distinct design motivation enabled by the workflow. All examples were designed and composed in MiXR, then 3D printed on a Stratasys J55 Prime using four Vero color resins (cyan, magenta, yellow, and white) together with WSS150 water-soluble support material. Print times ranged from roughly 4 to 12 hours depending on part size, and post-processing consisted only of submerging the part in water to dissolve the support. This is a comparatively forgiving fabrication process, and we do not expect these results to generalize to all printing modes (e.g., FDM printing without soluble supports, where overhangs and unsupported spans constrain what is printable).

\subsection{Personalized Fabrication}
\begin{figure}
    \centering
    \includegraphics[width=0.8\linewidth]{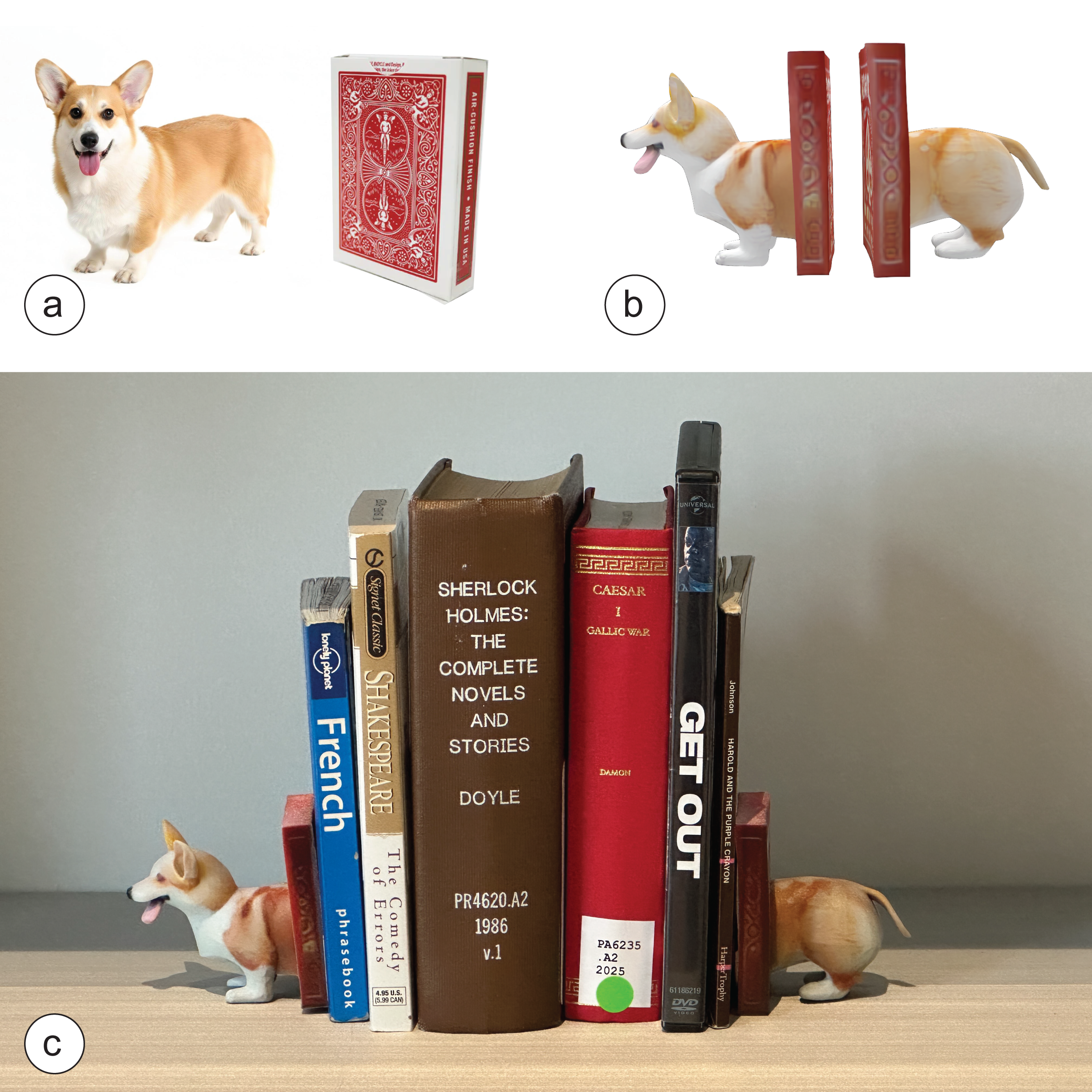}
   \caption{A personalized book support composed from a user's pet corgi. (a--b)~The dog model is split into halves and a deck of cards is placed between them as a support surface. (c)~The 3D printed result.}

    \label{fig:application_dog}
\end{figure}

MiXR enables users to turn personally meaningful objects into functional designs whose geometry could not be specified through a text prompt alone. In Figure~\ref{fig:application_dog}, a user creates a book support in the likeness of their pet corgi. The dog model is split into front and back halves via paint-based segmentation, and a captured deck of cards is positioned vertically between them as a structural support surface. The composed model is refined and 3D printed as a functional artifact that is personally meaningful to the user.

The two halves print as a matched pair. The front half, carrying the head and forelegs, stands at one end of a row of books and the rear half at the other, so the dog reads as passing behind the row (Figure~\ref{fig:application_dog}c). The card deck contributes the flat vertical faces that the books bear against, a surface the dog geometry alone does not provide, and the segmentation plane determines where each half meets that surface. What the harvested geometry contributes here is specificity rather than convenience: the proportions belong to this particular animal, which is the property that makes the object personally meaningful and the one a generic asset cannot supply.

\subsection{Creative Reuse}
\begin{figure}
    \centering
    \includegraphics[width=0.8\linewidth]{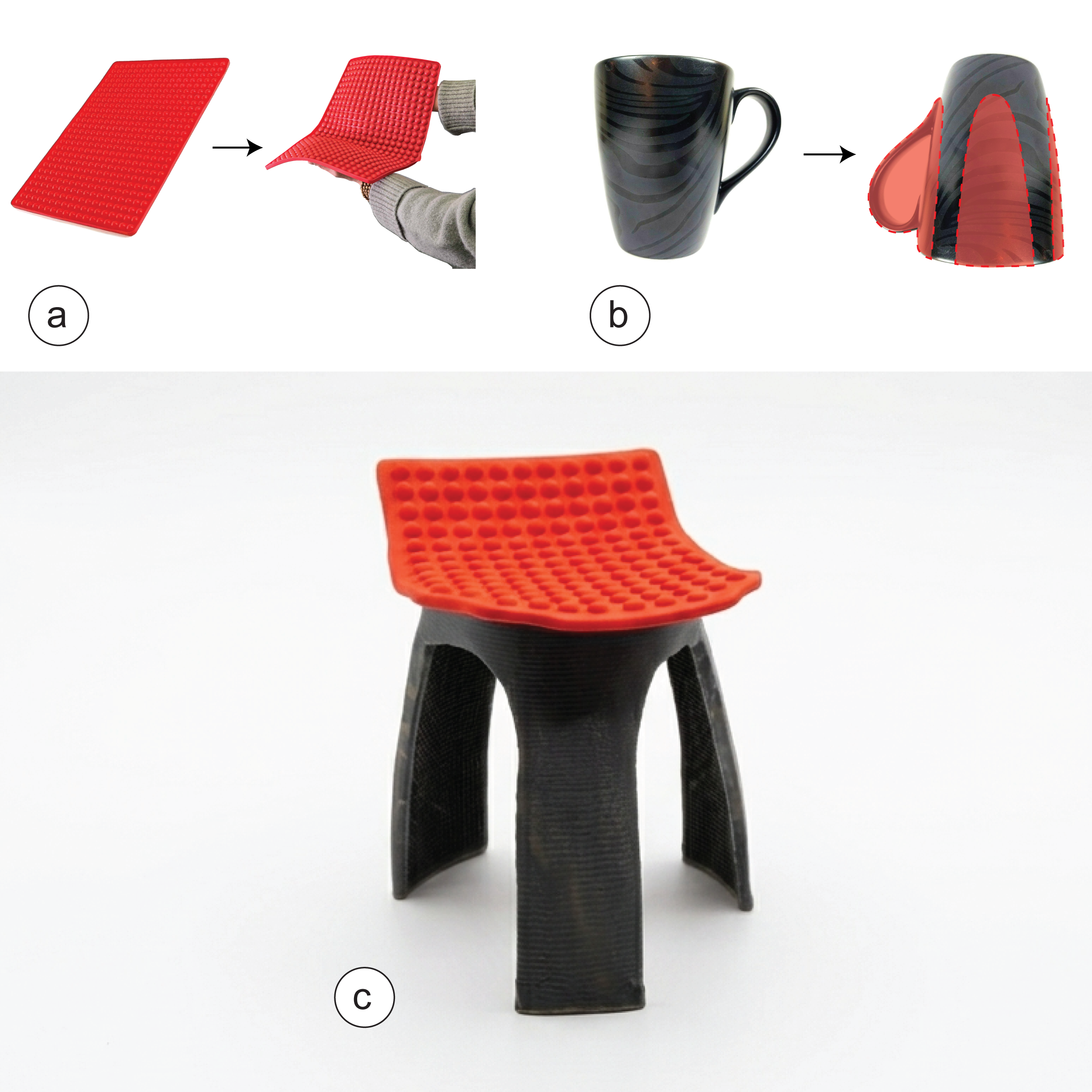}
    \caption{A chair designed from household objects. (a)~A non-slip mat is physically shaped into a seat and captured as a proxy mesh. (b)~A coffee mug is carved to extract leg geometry. (c)~The composed and 3D printed result.}

    \label{fig:application_chair}
\end{figure}

MiXR supports creative reuse by allowing users to repurpose everyday objects as geometric building blocks. In Figure~\ref{fig:application_chair}, a user designs a miniature chair from household materials. They physically bend a non-slip mat into the desired seat curvature and capture it as a proxy mesh. For the legs, they capture an upside-down coffee mug and selectively delete regions to extract the desired geometry. Neither object was designed as furniture, but MiXR's segment-and-compose workflow lets users reinterpret their affordances.

The printed stool keeps both sources legible: the seat retains the raised bump texture of the original mat, and the legs retain the flared profile of the inverted mug (Figure~\ref{fig:application_chair}c). The seat curvature is the notable input here, because it was specified by a physical gesture rather than a modeling operation. The user bent the mat until the curve looked right and captured that state, which is a form of geometric input with no direct equivalent in prompt-based or parametric workflows, where the same shape would have to be described or reconstructed through control points.

\subsection{Embodied Human-Centered Design}
\begin{figure}
    \centering
    \includegraphics[width=0.8\linewidth]{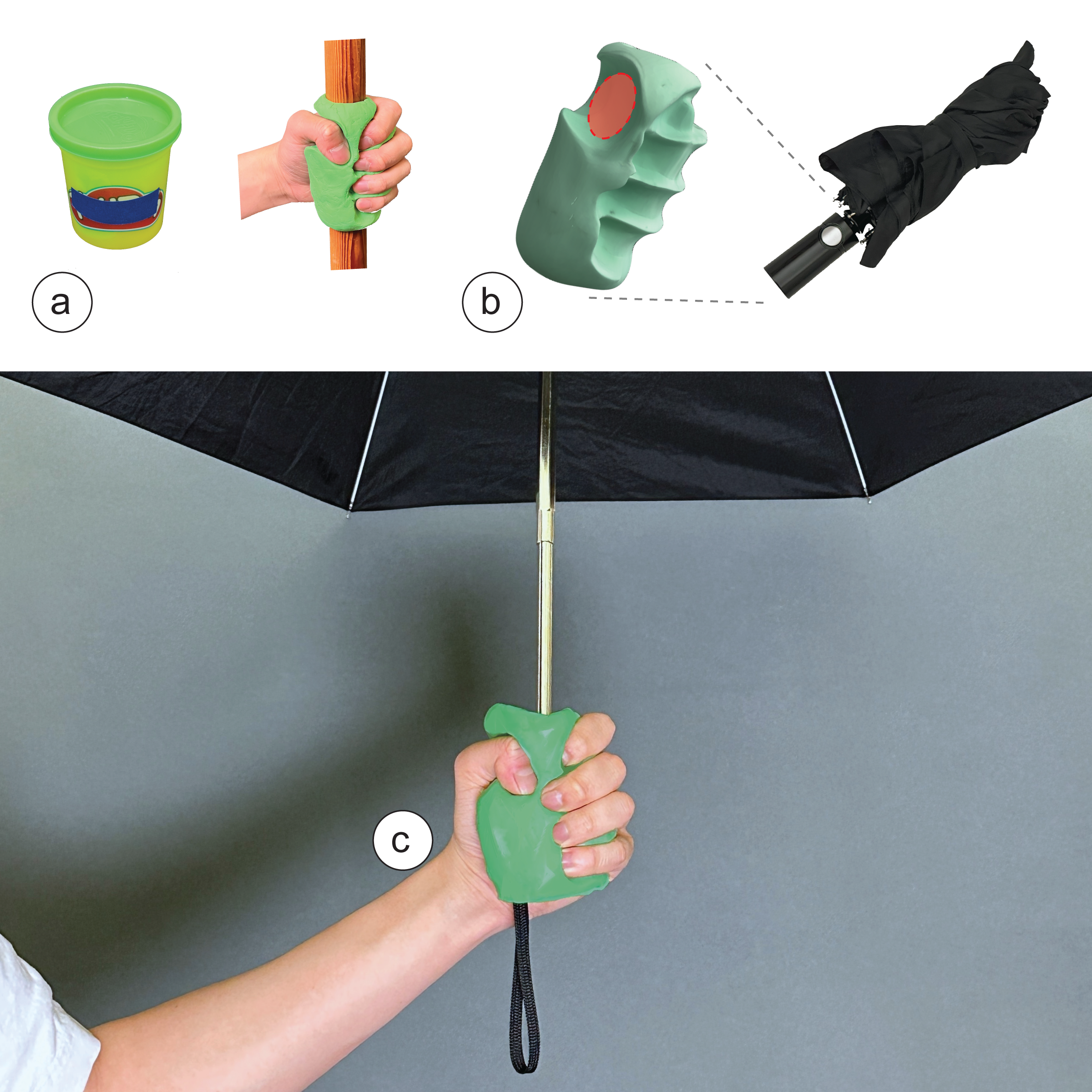}
\caption{A custom accessibility grip for an umbrella. (a)~Modeling dough is molded around a rod to capture the user's hand shape. (b)~The grip is composed over the umbrella handle with segmentation clearing the button area. (c)~The 3D printed assistive attachment.}

    \label{fig:applications_umbrella}
\end{figure}

MiXR also supports designs that must conform to both a user's body and a target object. In Figure~\ref{fig:applications_umbrella}, a user with limited grip strength molds modeling dough around a rod to capture a hand-conforming grip shape, then composes it over their umbrella's handle. Segmentation removes the portion obstructing the open/close button, a constraint that emerges only from inspecting the physical object in situ. The 3D printed grip attaches as an assistive device whose geometry is directly informed by the user's hand shape and the target object.

The printed part reproduces the finger indentations left in the dough, so the hand returns to the posture it was captured in, and the cleared region keeps the open/close button reachable while the grip is held (Figure~\ref{fig:applications_umbrella}c). This artifact resolves two independent physical references at once, the user's hand and this umbrella's handle diameter and button placement. Neither reference is available to a text prompt or an asset library, and both were captured in the environment where the object is used, which is the condition that in-situ composition is specifically suited to.

\section{User Study}
\label{sec:user_evaluation}
Our goal was to evaluate whether compositional manipulation of 3D segments gives users more precise control over the resulting model than delegating composition to an AI model through verbal instructions. We conducted a controlled user study addressing three questions:

\begin{enumerate}
    \item Does direct spatial manipulation of segments give users more precise control and produce models that more closely match a target design than verbal description?
    \item Does this advantage hold across both abstract and real-world tasks?
    \item Do users feel the additional effort of manual composition is justified by added design control?
\end{enumerate}

In the following subsections, we describe the study design and baseline condition, participant demographics, and the tasks and procedure used in the evaluation.

\subsection{Study Design \& Baseline Condition}

The study followed a within-subjects $2 \times 2$ factorial design with two independent variables: \textbf{Condition} (MiXR vs.\ Generative Composition) and \textbf{Task} (Building Blocks vs.\ Real-World Objects). Each participant completed all four combinations, yielding four experimental runs per participant. Both task order and condition order were fully counterbalanced using a Latin Square design, ensuring that each combination appeared equally often in each position. Each session lasted approximately 90~minutes.

\paragraph{Baseline: Generative Composition.}
As a baseline, we implemented a generative composition workflow within the same AR environment. Participants captured screenshots of the physical objects and verbally described the desired spatial arrangement. The screenshots and description were passed to Gemini~\cite{gemini3report2025}, which generated an image depicting the described composition, then converted to a 3D model via SAM3D~\cite{sam3d}. Participants could iterate by re-describing or re-taking screenshots.

The key distinction is the locus of control: in the baseline, spatial composition is delegated to a VLM~\cite{gemini3report2025} that interprets verbal intent, generates an image, which is then converted to a 3D model with SAM3D. In MiXR, the user directly manipulates segments in 3D space, and the generative model's role is limited to refining this user-defined arrangement into unified geometry. Both conditions used the same physical objects, target models, and XR headset, and both ultimately reconstruct geometry with SAM3D. They differ in more than interaction style, however: the baseline additionally routes composition through a text-to-image generation stage (Gemini) that MiXR does not use. We therefore treat this study as a comparison between two complete composition workflows, direct segment-level composition versus VLM-mediated generative composition, rather than an isolated test of interaction modality alone.

\subsection{Participants}
We recruited 12 participants (7 male, 5 female), aged 19--33 years ($M = 23.1$, $SD = 3.7$), through institute mailing lists and snowball sampling. All participants were novices with no professional 3D modeling experience; prior experience with 3D modeling tools was low ($M = 3.17$, $SD = 1.59$, on a 7-point scale from ``Beginner'' (1) to ``Expert'' (7). Familiarity with AR/VR headsets was similarly low ($M = 2.42$, $SD = 1.38$), while familiarity with AI generation tools was moderate ($M = 4.17$, $SD = 1.53$). Self-reported comfort with hands-on spatial activities (crafts, building, puzzles) was moderate to high ($M = 5.17$, $SD = 1.27$, on a 7-point scale). Participants were compensated \$30 for their time.

\subsection{Tasks and Procedure}
Each participant completed four experimental runs across two tasks, one run per cell of the $2 \times 2$ design. Both tasks required participants to \emph{recreate a target model} as closely as possible using the assigned condition in under 10 minutes.

\paragraph{Target Models.}
\begin{figure}
    \centering
    \includegraphics[width=\linewidth]{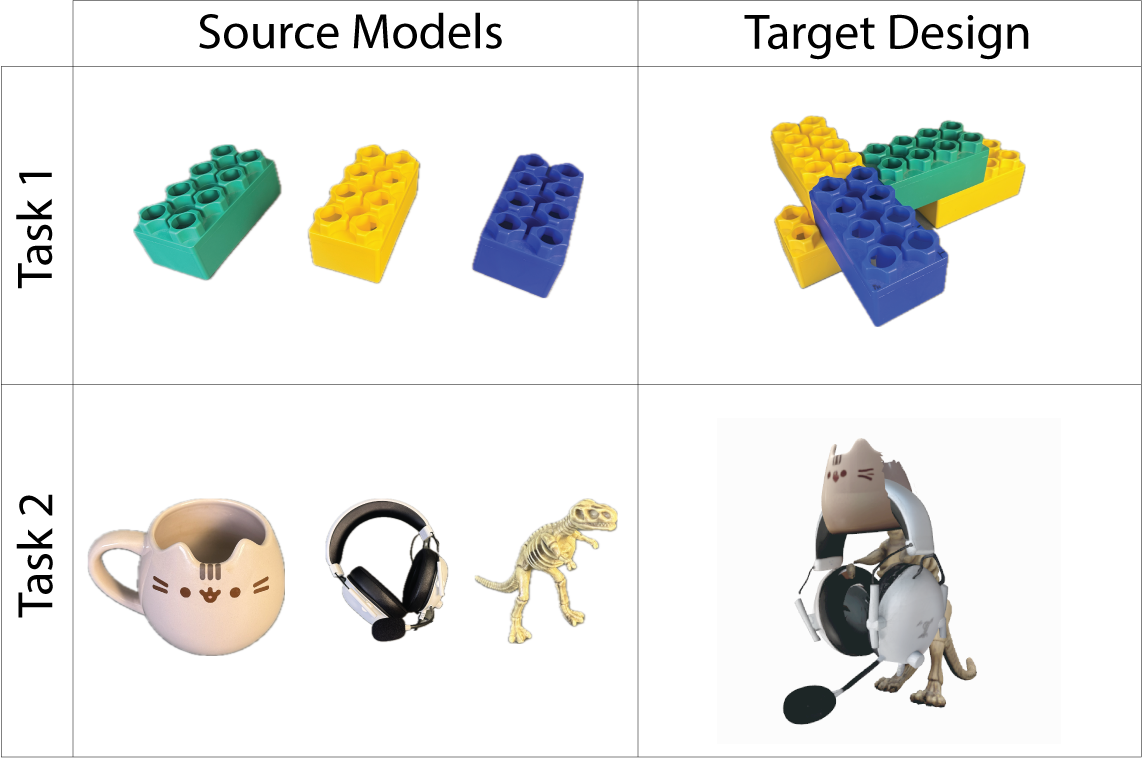}
    \caption{Source objects and target designs for the two study tasks. Task~1 uses three interlocking building blocks with regular geometry. Task~2 uses real-world objects with organic shapes (a cat mug, headphones, and a dinosaur figurine) composed into a novel form.}
    \label{fig:user_study-source-target_models}
\end{figure}

The two tasks were designed to vary the nature of the source material and the spatial reasoning required (Figure~\ref{fig:user_study-source-target_models}). Task~1 used three interlocking building blocks of the same dimension but different colors, arranged into a specific configuration. Task~2 required participants to compose a target design from segments of real-world everyday objects: a cat-shaped mug, a pair of headphones, and a dinosaur skeleton, combining them into a unified novel form. For both tasks, participants were shown the target 3D model before starting but were not permitted to use it as input to any generative model. By pairing an abstract building block arrangement with a freeform real-world object composition, we test whether the two conditions (MiXR and Generative Composition) exhibit any meaningful differences across qualitatively different spatial reasoning demands.

\paragraph{Procedure.}
Each session began with a pre-study questionnaire capturing demographics and prior experience, followed by a 10-minute tutorial covering both conditions using a practice object not in the study set.

Participants then completed the four experimental runs in counterbalanced order. For each run, they reviewed the reference 3D model, then completed the recreation task under the assigned condition within a 10-minute time limit. After each run, they completed a NASA TLX questionnaire and a post-task questionnaire with 10 Likert items (7-point scale) across three constructs: \emph{Task Outcome}, \emph{Effort \& Enjoyment}, and \emph{Tool Responsiveness}.

After all four runs, participants completed forced-choice comparisons between conditions (overall preference, control, accuracy, effort, confidence, enjoyment, and spatial arrangement), followed by a semi-structured interview on creative strategies, sense of control, and attachment to designs.

\subsection{Data Analysis}

We analyze the NASA TLX scores and post-task questionnaires using two-way repeated-measures ANOVAs with factors Condition (MiXR vs.\ Generative Composition) and Task (Building Blocks vs.\ Real-World Objects). To compare the two conditions overall, we report each participant's scores across both tasks using paired $t$-tests with Holm--Bonferroni correction for multiple comparisons. Post-study forced-choice preferences are analyzed with binomial tests. Qualitative data from open-ended questionnaire items and semi-structured interviews were analyzed using Reflexive Thematic Analysis~\cite{braun2006using} to extract high-level themes.

 \begin{figure}
    \centering
    \includegraphics[width=\linewidth]{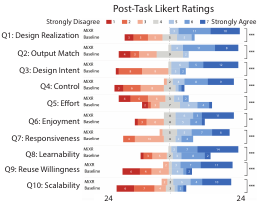}
    \caption{Post-task Likert ratings by condition (7-point scale; $N=12$). Each bar shows the distribution of individual responses. MiXR was rated significantly higher on all 10 items after Holm--Bonferroni correction. (Significance levels: {*}~$p<.05$, {**}~$p<.01$, {***}~$p<.001$.)}

    \label{fig:user_study_post_task}
\end{figure}

 \begin{figure}
    \centering
    \includegraphics[width=0.8\linewidth]{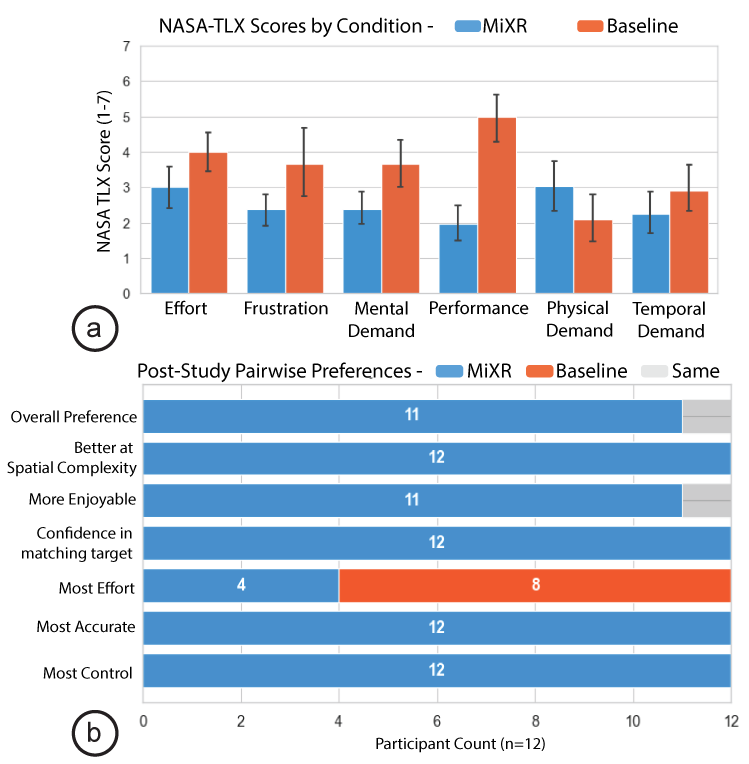}
    \caption{(a)~NASA TLX subscale ratings by condition (7-point scale; lower is better (except Performance, which is reverse-coded). (b)~Post-study forced-choice preferences ($N=12$). MiXR was unanimously preferred on control, accuracy, confidence, and spatial complexity. Error bars show $\pm 1$ SE.}

    \label{fig:user_study_nasa_choice}
\end{figure}

Figures~\ref{fig:user_study_post_task} and~\ref{fig:user_study_nasa_choice} summarize the quantitative results. We report the user-reported differences between the two conditions, then present qualitative themes from open-ended responses and interviews.

\subsection{Quantitative Findings}
\label{sec:results}

\paragraph{Post-Task Ratings.}
MiXR was rated significantly higher than the Generative Composition baseline on all 10 Likert items after Holm--Bonferroni correction (Figure~\ref{fig:user_study_post_task}). The largest differences were on task outcome items: participants more strongly agreed that the output matched the target (Q2: MiXR $M=6.08$ vs.\ Baseline $M=3.08$; $p_\text{holm}<.001$), that they could create the design they had in mind (Q1: $6.29$ vs.\ $3.29$; $p_\text{holm}<.001$), and that they felt in control (Q4: $5.92$ vs.\ $3.04$; $p_\text{holm}<.001$). Tool responsiveness items followed the same pattern: MiXR was rated higher on matching expectations (Q7: $5.50$ vs.\ $3.67$; $p_\text{holm}<.001$), learnability (Q8: $6.42$ vs.\ $4.46$; $p_\text{holm}=.002$), and perceived scalability (Q10: $5.75$ vs.\ $2.83$; $p_\text{holm}<.001$). Participants also found MiXR more enjoyable (Q6: $5.88$ vs.\ $4.04$; $p_\text{holm}=.007$), reporting that the Baseline required more effort (Q5: Baseline $M=4.25$ vs.\ MiXR $M=3.04$; $p_\text{holm}=.023$).

 \paragraph{Workload.}
NASA TLX scores were significantly lower for MiXR overall (MiXR $M=2.50$ vs.\ Baseline $M=3.56$; $p<.01$; Figure~\ref{fig:user_study_nasa_choice}a). The largest subscale difference was Performance, where participants rated themselves  more successful with MiXR ($1.96$ vs.\ $5.00$; $p_\text{holm}<.001$). Mental Demand was also significantly higher for the Baseline ($3.67$ vs.\ $2.38$; $p_\text{holm}=.020$), reflecting the cognitive load of translating spatial intent into language. Physical Demand was the one dimension where MiXR trended higher ($3.04$ vs.\ $2.08$), consistent with its hands-on interaction, but the results were not significant ( $p > 0.05$).

\paragraph{Forced-Choice Preferences.}
Post-study comparisons were near-unanimous (Figure~\ref{fig:user_study_nasa_choice}b). Participants unanimously selected MiXR for more control, more accurate, high confidence in matching the target, and better for spatial complexity (all 12/12, $p<.001$). MiXR was also preferred overall (11/12, $p<.001$) and rated as more enjoyable (11/12, $p<.001$). For effort, 8 of 12 attributed more effort to the Baseline ($p=.388$, n.s.), consistent with the TLX pattern. Appendix~\ref{app:user_study} (Figs.~\ref{fig:user_lego} and~\ref{fig:user_headphones}) shows the complete set of participant outputs for both conditions and both tasks, illustrating this gap directly across all 12 participants rather than in aggregate.

 \subsection{Qualitative Findings}

\paragraph{Direct manipulation supports intuitive spatial reasoning.}
Participants consistently contrasted the directness of MiXR with the indirectness of verbal description. P10 captured the core distinction: ``\textit{for direct manipulation, I could just put the parts relative to another versus for verbal I would have to think about how to describe the spatial relationships.}'' P8 described verbal description as disorienting: ``\textit{verbal description was all over the place---I kept jumping back-and-forth between ideas, but approach~A was just like I dealt with each component till I was ready to move on.}''

\paragraph{Effort as engagement vs. effort as struggle.}
Participants experienced MiXR's physical effort as engaging rather than taxing: ``\textit{with direct manipulation I felt really engaged\ldots it seems like [the time] flew by super fast, and with the AI one I just tried to prompt it a few times doing the exact same thing and it just got boring}'' (P3). P9 summarized: ``\textit{I had more control and had to work more physically, though less mentally, so I got less frustrated.}''

\paragraph{Ownership and attachment.}
11 of 12 participants felt more attached to designs composed through direct manipulation, attributing this to personal investment: ``\textit{the ones I generated from parts were literally moved piece by piece by my own gestures, so that process feels more tangible in memory}'' (P1), and authorship: ``\textit{it felt like work that I was doing, as opposed to just telling someone else what to do}'' (P10).

\paragraph{Participants envision hybrid workflows.}
Nearly all participants independently proposed combining both approaches: generative composition for a coarse initial pass, then direct manipulation for refinement. P5 described: ``\textit{verbally get a draft assembly, but still have separate objects you can manipulate}''. When asked whether they would still want interactive composition even if AI could perfectly execute a verbal description, participants were split (6~yes, 4~no, 2~depends), but even those who said no wanted post-hoc editing: ``\textit{I would hope that I'm able to make edits to parts of the model rather than the whole thing}'' (P3).

\section{Discussion and Limitations}
 In this section, we distill design guidelines from our findings, discuss the hybrid workflow that participants envisioned, identify limitations of the current system and evaluation, and discuss future directions.

\subsection{Design Guidelines for Balancing User Control and AI Delegation}

Drawing on our results, we distill three design guidelines for compositional 3D tools in XR.

\paragraph{Treat user-defined spatial arrangement as a hard constraint.}
Participants were consistently frustrated when the baseline model reinterpreted their specified arrangement. MiXR avoids this by preserving the user's composition in the latent representation and restricting the model's role to geometric integration.

\paragraph{Default to direct manipulation for spatial composition, regardless of geometric complexity.}
Our study indicates that MiXR's advantage held across both regular and organic geometry. Even for Task~1's axis-aligned arrangements, verbal description performed poorly, suggesting the bottleneck is not geometric complexity but the fundamental difficulty of translating spatial intent into language with sufficient precision.

\paragraph{Design for iterative refinement rather than single-shot composition.}
Participants routinely returned to Segment and Compose modes to adjust placements and refine selections. Compositional workflows should surface incremental feedback, such as per-segment previews prior to full decoding, to reduce the cost of late-stage error discovery.

\subsection{The Hybrid Workflow as a Design Pattern}

Ten of twelve participants independently proposed combining verbal input for object generation with direct manipulation for spatial arrangement. This convergence suggests an unmet need, consistent with what the study demonstrates: verbal description is effective for specifying \emph{what} to generate, while direct manipulation is necessary for specifying \emph{where} and \emph{how} components are arranged. The two modalities are complementary rather than substitutable, and MiXR's architecture already reflects this division. Future systems can further explore this system-design space, leveraging language for content instantiation and reserving embodied interaction for spatial composition and refinement.

\subsection{Limitations and Future Work}
Our system design and study have several limitations that suggest directions for future work. Firstly, we focused our evaluation on a single generative model, SAM3D, due to its state-of-the-art reconstruction fidelity, generalization across inputs, two-stage generative pipeline, and capacity of taking text and image input together. We plan to open-source our implementation to support future analysis across additional generative models as they continue to evolve. Secondly, MiXR captures objects as static meshes with no representation of moving parts or material compliance, limiting the range of functional artifacts users can design. Thirdly, because composition happens in-situ in AR, MiXR provides an embodied sense of scale: users size and place geometry directly against the physical objects and bodies it must fit, as in the umbrella grip's custom fit to the user's hand (Sec.~\ref{sec:applications}). This embodied scale matching is not a substitute for precision dimensional constraints, however; the system has no mechanism to enforce that two parts mate to a specific numeric tolerance, such as a machined interface or screw thread, making it unsuitable for fabrication tasks that require exact fit rather than embodied scale matching. Future work can explore measurement tools within the generative workflow to support these precision cases. Composition fidelity is bounded by the decoder's fixed $64^3$ voxel grid, which degrades with large spatial separations and scale mismatches (Sec.~\ref{sec:technical}). MiXR also inherits the reconstruction quality of SAM3D's single-image backbone, with no independent mechanism to correct proxy mesh errors. Because MiXR's repair mechanism operates through the backbone model's own decoder rather than a bespoke geometric heuristic, however, its output quality should improve as off-the-shelf 3D generative models improve, without requiring changes to MiXR itself. Finally, the study recruited 12 novice participants from a university setting; results may not generalize to professional designers, or users with physical limitations affecting hand tracking in AR.

\section{Conclusion}

We presented MiXR, an AR system that harvests geometry from real-world objects and composes it into unified, fabricable meshes. Users capture physical objects as 3D proxy meshes, extract segments through paint-based selection, and arrange them explicitly in space; a generative model, intervening on SAM3D's structured latent representation, then integrates the composition into unified geometry, repairing the open boundaries that segmentation introduces. Two technical evaluations characterize this pipeline: a controlled transform sweep establishes the operating regime in which composition remains reliable, and a 100-sample watertightness stress test shows the decode stage converts open, non-manifold segmented input into watertight and manifold meshes in 86\% of tested cases. We demonstrated the pipeline end-to-end on three fabricated, functional artifacts built from dimensionally-specific, harvested geometry that general-purpose asset libraries cannot provide. A controlled user study further shows that novices can reliably drive this pipeline: participants rated MiXR designs as more closely matching their intended target and reported greater perceived control than with a VLM-mediated generative composition baseline. MiXR's scope is bounded by the decoder's fixed-resolution latent grid and by the absence of precision dimensional constraints: its in-situ workflow provides embodied scale matching against physical objects and bodies, but not the exact numeric tolerances that mating or machined parts require, marking the boundary between MiXR's compositional prototyping role and the precision fabrication tasks that remain the province of conventional CAD.

\section*{Disclosure of Generative AI Use}

Large language model assistance was used in preparing this paper for drafting and revising portions of the manuscript text, and for generating and revising portions of the evaluation and analysis code. All experimental design, data collection, and interpretation of results were performed by the authors. The authors verified the reported results against the underlying data and take full responsibility for the content of this paper.

\bibliographystyle{ACM-Reference-Format}
\bibliography{references}

\pagebreak

\clearpage

\appendix

\section{Technical Evaluation: Per-Pair Breakdowns}
\label{app:tech_eval}

Section~\ref{sec:technical} reports aggregate Mesh IoU and Chamfer Distance
across all 25 ordered category pairs.
This section provides the per-pair heatmaps for each transform sweep
(translation, rotation, and scale), as well as the extended translation
sweep (0--10 object extents) referenced in Section~\ref{sec:technical}.
Because the batch evaluation samples each sweep on a discrete grid,
the appendix heatmaps use the nearest available sampled step to the
paper-facing target value rather than an interpolated intermediate value;
the values below are rounded for readability.

\subsection{Extended Translation Sweep (0--10 Object Extents)}
\label{app:translation_extended}

\begin{figure}[h]
  \centering
  \includegraphics[width=\linewidth]{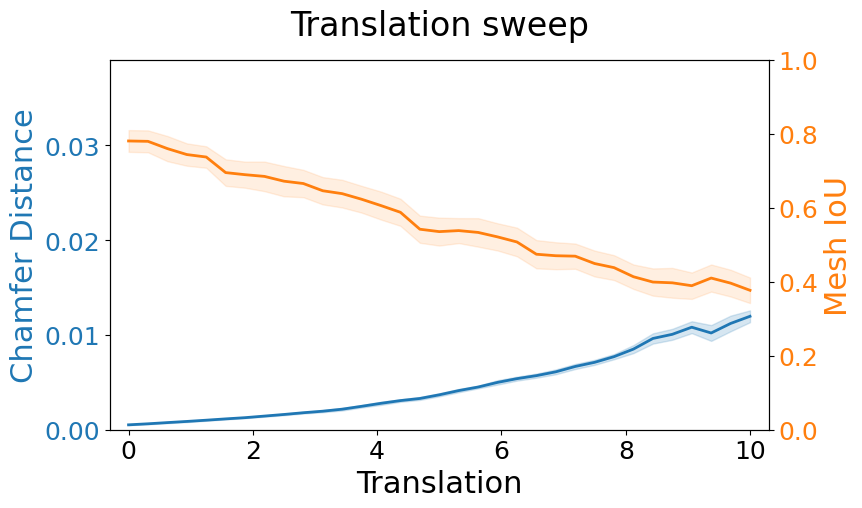}
  \caption{Mean Chamfer Distance (squared) and mean Mesh IoU across all
           25 ordered category pairs for the extended translation sweep
           (0--10 object extents). Shaded bands show 95\% confidence intervals.
           Degradation accelerates beyond ${\sim}$5 extents; large-separation
           values represent configurations that are unlikely in practice, as
           users composing coherent designs will keep components in close
           proximity.}
  \label{fig:sweep_translation_10}
\end{figure}

Figure~\ref{fig:sweep_translation_10} reports the full translation sweep
extended to 10 object extents, as referenced in Section~\ref{sec:technical}.
Beyond the practical composition regime ($\lesssim$2.5 extents), fidelity
continues to degrade for all pairs, confirming that the monotonic trend shown
in Figure~\ref{fig:tech-eval-sweeps} of the main paper persists at larger separations.

\subsection{Translation Heatmap at 1.88 Additional Object Extents}
\label{app:heatmap_translation}

\begin{figure*}[h]
  \centering
  \includegraphics[width=\linewidth]{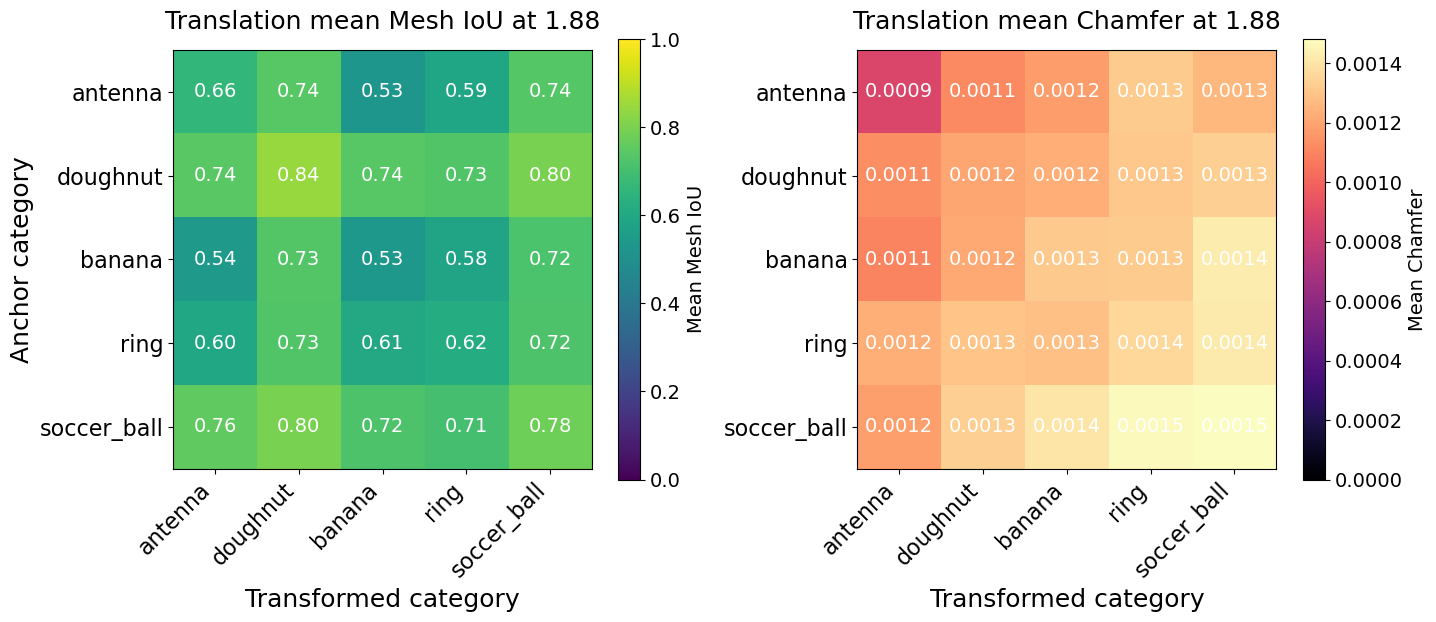}
  \caption{Per-pair heatmaps for translation at 1.88 additional object
           extents. The left panel shows Mesh IoU and the right panel shows
           squared Chamfer Distance for each ordered anchor--target pair.
           At this moderate separation, degradation is already visible but
           remains meaningfully pair-dependent, with compact convex geometries
           such as soccer ball--doughnut remaining comparatively robust.}
  \label{fig:heatmap_translation}
\end{figure*}

Figure~\ref{fig:heatmap_translation} shows per-pair Mesh IoU and Chamfer
Distance at a fixed translation offset of 1.88 additional object extents
beyond the built-in baseline separation. This is the nearest sampled step in
the 33-step linear translation grid to the nominal paper-facing target of
2.00 additional extents.
Each cell corresponds to one ordered anchor--target pair drawn from the five
Objaverse categories (antenna, banana, doughnut, ring, soccer ball).

\subsection{Rotation Heatmap at $33.8^\circ$}
\label{app:heatmap_rotation}

\begin{figure*}[h]
  \centering
  \includegraphics[width=\linewidth]{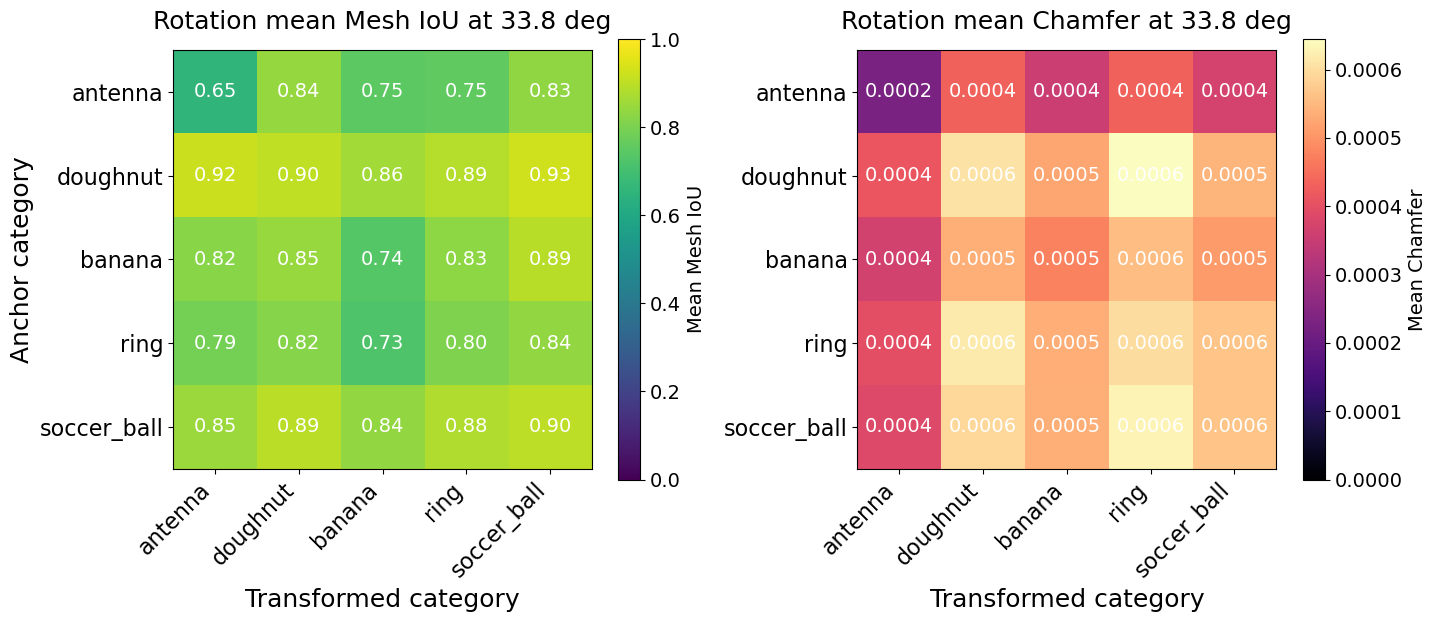}
  \caption{Per-pair heatmaps for rotation at $33.8^\circ$. The left panel shows
           Mesh IoU and the right panel shows squared Chamfer Distance.
           Most ordered pairs remain close to their baseline values, consistent
           with the aggregate finding that rotation is comparatively benign.}
  \label{fig:heatmap_rotation}
\end{figure*}

Figure~\ref{fig:heatmap_rotation} shows per-pair results for the rotation
sweep at a fixed relative rotation of $33.8^\circ$. This is the nearest
sampled step in the 33-step linear rotation grid to the nominal paper-facing
target of $30^\circ$. Consistent with the
aggregate reported in Section~\ref{sec:technical}, values remain tightly
clustered across the 25 ordered pairs, confirming that modest relative
reorientation does not meaningfully expand the spatial support of the composed
SLAT volume.

\subsection{Scale Heatmap at $1.54\times$}
\label{app:heatmap_scale}

\begin{figure*}[h]
  \centering
  \includegraphics[width=\linewidth]{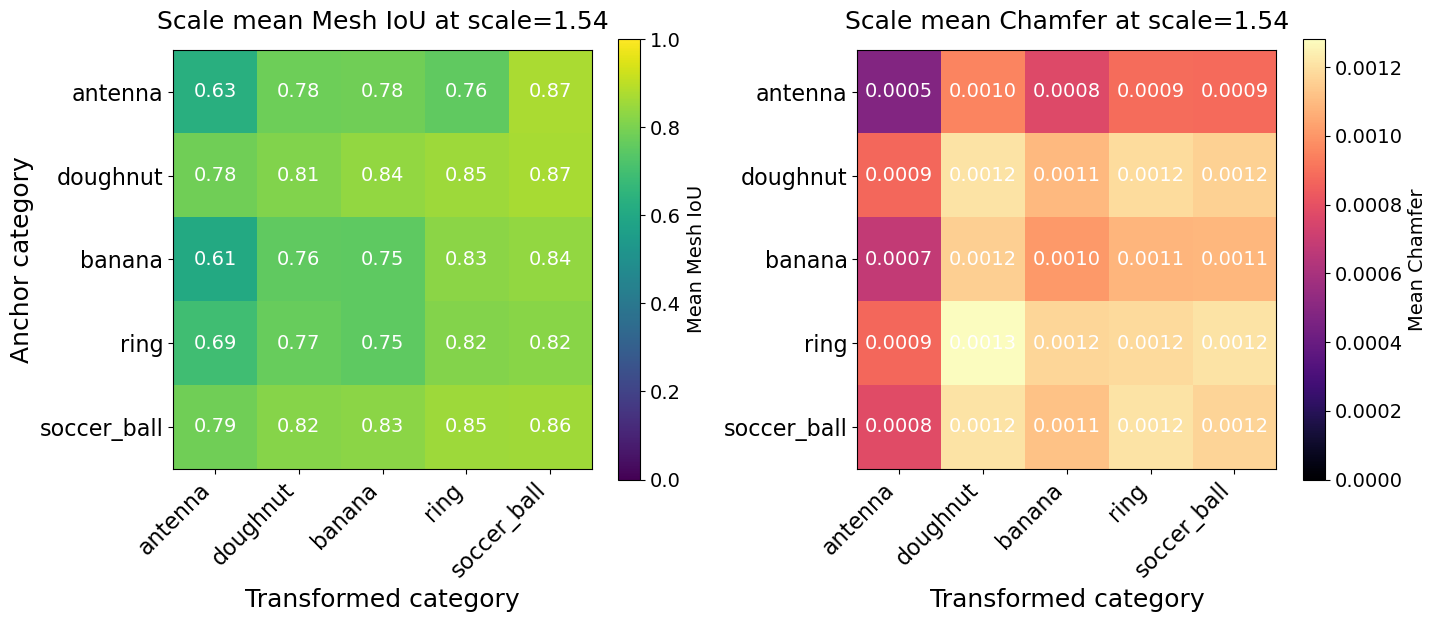}
  \caption{Per-pair heatmaps for uniform scale at
           $1.54\times$. The left panel shows Mesh IoU and the
           right panel shows squared Chamfer Distance. At this modest
           enlargement, most pairs remain close to the stable regime, while
           elongated or asymmetric geometries begin to show the early signs of
           divergence later amplified at larger scale factors.}
  \label{fig:heatmap_scale}
\end{figure*}

Figure~\ref{fig:heatmap_scale} shows per-pair results for the uniform scale
sweep at a fixed scale factor of $1.54\times$. This is the nearest
sampled step in the 33-step geometric scale grid to the nominal paper-facing
target of $1.50\times$. This slice lies just beyond
the identity regime and provides a qualitative view of how pairs begin to
separate before the much sharper nonlinear Chamfer rise that emerges at larger
scale factors.

\section{User Study: Full Participant Outputs}
\label{app:user_study}

This section presents the complete set of participant outputs (P1--P12) for
both conditions (Baseline and MiXR) across both study tasks.
Source objects and target designs are shown at the top of each figure for
reference. All outputs were exported directly from the AR system at the end
of each 10-minute trial.

\subsection{Task 1 --- Building Block Arrangement}
\label{app:results_lego}

\begin{figure*}[h]
  \centering
  \includegraphics[width=\linewidth]{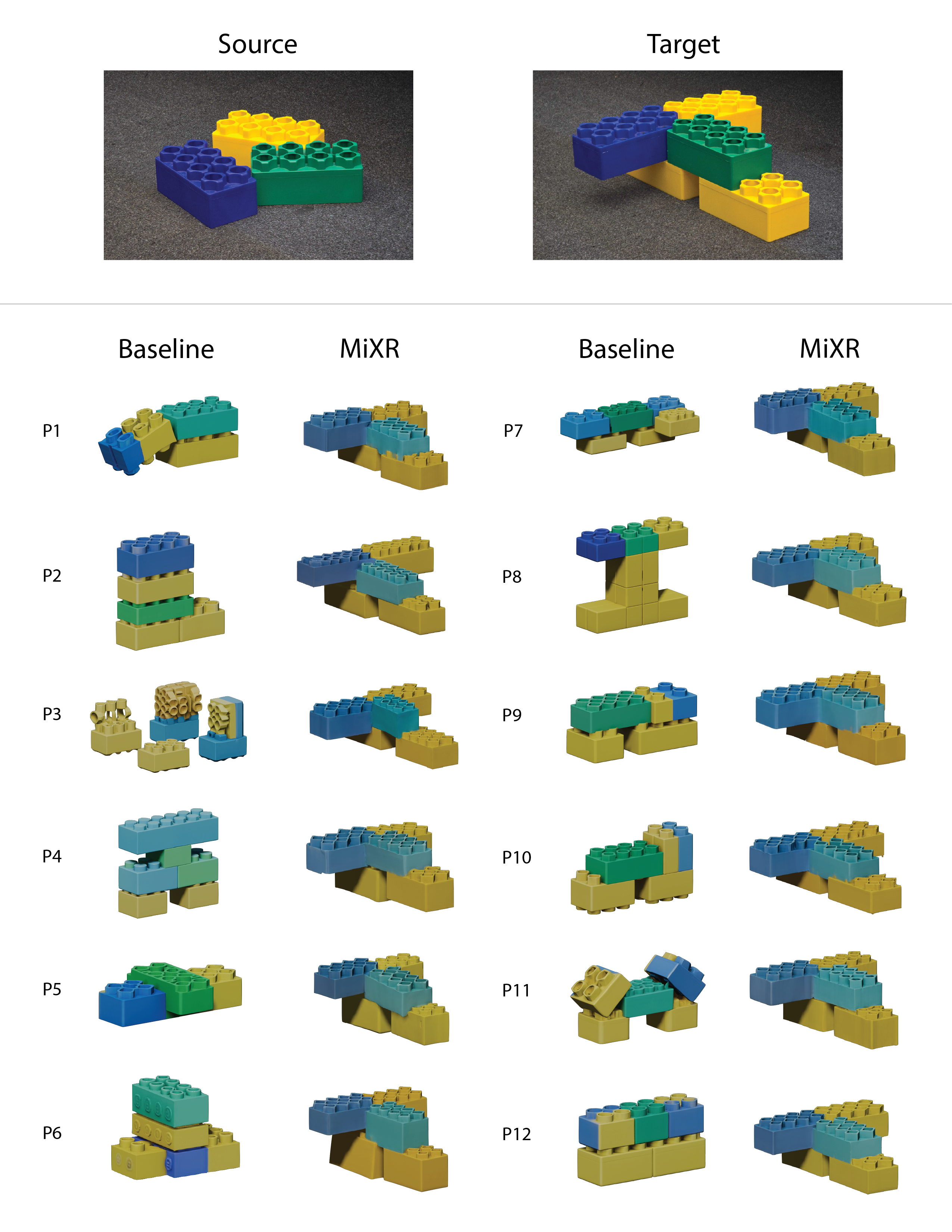}
  \caption{All 12 participant outputs for Task~1 (building block arrangement).
           Source and target designs are shown at the top.
           For each participant (P1--P12), the Baseline output is shown on
           the left and the MiXR output on the right.}
  \label{fig:user_lego}
\end{figure*}

Figure~\ref{fig:user_lego} shows all participant outputs for Task~1, in which
participants recreated a specific spatial arrangement of three interlocking
building blocks (blue, green, and yellow) of equal dimensions.
MiXR outputs broadly preserve the target's layered structure across
participants, whereas Baseline outputs diverge in block placement,
relative orientation, and colour assignment.

\subsection{Task 2 --- Real-World Object Composition}
\label{app:results_headphones}

\begin{figure*}[h]
  \centering
  \includegraphics[width=\linewidth]{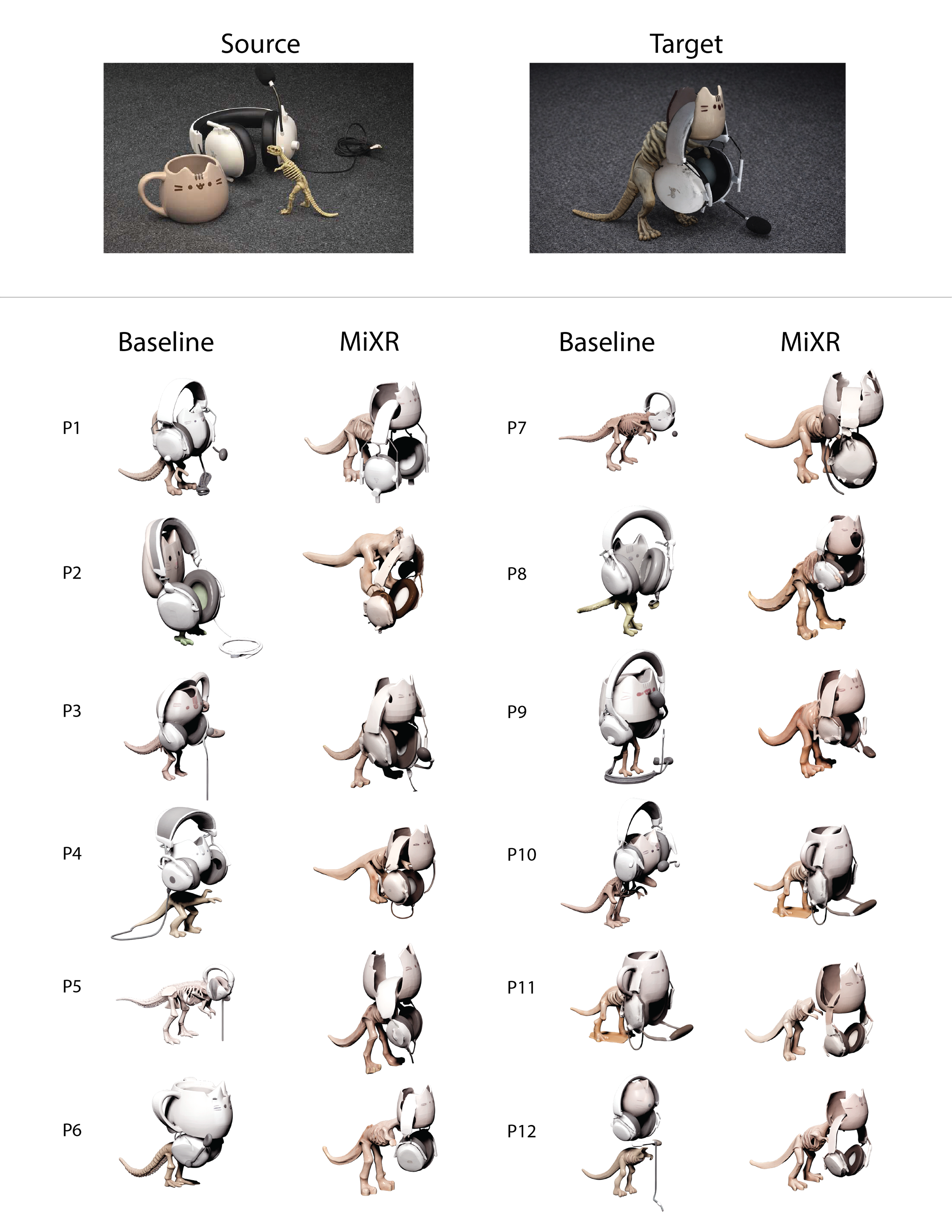}
  \caption{All 12 participant outputs for Task~2 (real-world object
           composition). Source and target designs are shown at the top.
           For each participant (P1--P12), the Baseline output is shown on
           the left and the MiXR output on the right.}
  \label{fig:user_headphones}
\end{figure*}

Figure~\ref{fig:user_headphones} shows all participant outputs for Task~2,
in which participants composed segments of a cat-shaped mug, a pair of
gaming headphones, and a dinosaur skeleton into a target novel form.
MiXR outputs consistently preserve the dinosaur body as the primary
structural element with headphone components correctly positioned, while
Baseline outputs show high variability in part placement and relative scale.

\end{document}